\documentclass[journal]{IEEEtran}
\usepackage{psfig,amssymb,amsmath,amsfonts,amsthm}
\usepackage{multirow}

\newtheorem{theorem}{\textbf{Theorem}}
\newtheorem{lemma}{\textbf{Lemma}}
\newtheorem{corollary}{Corollary}
\newtheorem{definition}{Definition}

\author{Mohammad~Hamed~Firooz,~\IEEEmembership{Member,~IEEE,}
        Sumit~Roy,~\IEEEmembership{Fellow,~IEEE,}
\thanks{Authors are with the Department
of Electrical Engineering, University of Washington, Seattle,
WA, 98195 USA. e-mail: \{firooz,sroy\}@u.washington.edu

This work was supported in part from AFOSR under Grant FA9550-09-1-0298
}
}

\def\mathRV#1{\boldsymbol{\mathcal{ #1}}}

\title{Link Delay Estimation via Expander Graphs}

\begin{document}
\maketitle

\begin{abstract}
One of the purposes of network tomography  is to infer the status of parameters (e.g., delay) for the links inside a network through {\em end-to-end} probing between (external) boundary nodes along predetermined routes. In this work, we apply concepts from compressed sensing and expander graphs to the delay estimation problem. We first show that a relative majority of network topologies are {\em not} expanders for existing expansion criteria. Motivated by this challenge, we then relax such criteria, enabling us to acquire simulation evidence that link delays can be estimated for 30\% more networks. That is, our relaxation expands the list of identifiable networks with bounded estimation error by 30\%.   We conduct a simulation performance analysis of delay estimation and congestion detection on the basis of $l_1$ minimization, demonstrating that accurate estimation is feasible for an increasing proportion of networks.
\end{abstract}


\section{Introduction}\label{S:Intro}

Monitoring of link properties (delay, loss rates, etc.) in networks continues to be an integral requirement within any network management framework as part of monitoring its utilization and performance. The need for accurate and fast monitoring schemes has escalated in recent years due to the increasing popularity of new resource-consuming services (such as video-conferencing, Voice over IP, and online games) that require quality-of-service (QoS) guarantees \cite{eriksson2010toward,firooz2007multi}. The primary objective of this paper is to demonstrate how compressed sensing ideas may be applied to derive a fast delay monitoring algorithm that outperforms other schemes.

The term \emph{network tomography} was used in \cite{vardi1996nte} to encompass a class of approaches that seek to infer the internal link status from end-to-end measurements \cite{lawrence2006network,yu2000time}. A useful classification of network tomography methods for our purposes is as follows \cite{lawrence1001network}:

\begin{itemize}
\item \textbf{Cooperative Internal Nodes:} This method assumes that internal nodes on probe routes respond to {\em control} packets. For example, active tools such as a ping or a trace route, measure and report attributes of the round-trip path from a sender to the internal node based on probe packets \cite{richard94}. Beside complexity, the challenges of such methods arise from the fact that service providers do not own the entire network that is being probed and hence do not have access to the desired internal nodes for appropriate configuration \cite{bejerano2003rml,bejerano2006robust, ronasi2007enhanced}.

\item \textbf{End-to-End:}
In networks with a defined \emph{boundary}, it is assumed that access is available to nodes at the edge (but not to any in the interior). A boundary node sends probes to all (or a subset of) other boundary nodes to measure packet attributes on the path between network end points. These edge-based methods do not require exchanging control messages with any interior nodes. The primary challenge confronting such end-to-end probe-based link status estimation is that of identifiability, as discussed below \cite{caceres1999mbi,tsang2001pnt,eriksson2010toward}.
\end{itemize}
As the networks evolve toward more decentralized, uncooperative, and heterogeneous administrative (sub)domains, the availability of cooperative interior nodes is increasingly limited. Hence, end-to-end network diagnostic tools attract increasing attention.

Designing an end-to-end network tomography protocol can be based on two different considerations: (1) how to create a routing matrix $\mathbf{R}$ so that it covers the entire network; and (2) considering the routing matrix $\mathbf{R}$, how to estimate the desired attributes (e.g., link delay) of the network. The first issue is referred to in the literature as network coverage; two well-known algorithms are those that use multiple graphs and solve integer programming optimization \cite{bu2002ntg,rabbat2004multiple,hamed10}. In this paper, we focus on the second issue with the assumption that the routing matrix of the network is already given (a situation that holds in many practical examples). In our simulation, we use the shortest path as one of the possibilities for creating the routing matrix $\mathbf{R}$, but any other method is acceptable. Varying the type of method used does not affect the effectiveness of the algorithm.

For parameters such as delay, an additive linear model adequately represents the relationship between a measured path and an individual link delay \cite{coates2002it, bu2002ntg}, i.e.,
\begin{equation}
\mathbf{y} = \mathbf{R}\mathbf{x},
\label{E: basic tomography_stoch}
\end{equation}
where $\mathbf{x}$ is the $n \times 1 $ (unknown) vector of the individual link mean delay. The $ r \, \times \, n $ {\em binary} matrix $\mathbf{R}$ is the routing matrix for the network graph corresponding to the paths chosen for the probes (each row of the matrix correspond to a path), and $\mathbf{y} \; \in \; \mathbb {R}^r$ is the measured $r$-vector of end-to-end path delays. Although the focus of this paper is link delay, our approach readily applies to any other link attributes (such as $\log$ of packet loss rate), allowing such a linear relationship with end-to-end measurements.

In Eq. \eqref{E: basic tomography_stoch}, usually, the number of observations $r$ is much less than the number of variables $n$ (i.e., $r \; \ll \, n$) because the number of accessible boundary nodes is much smaller than the number of links inside the network. Thus, the number of variables in Eq. \eqref{E: basic tomography_stoch} to be estimated is much larger than the number of equations \cite{Castros2004nt}, leading to the generic nonuniqueness of solutions to Eq. \eqref{E: basic tomography_stoch}, i.e., the inability to uniquely determine link delay from end-to-end measurements \cite{xia2006inference}.
However, the problem of identifying only the (few) links with large delays (also known as congested links) suggests the possibility of improved mechanisms to solve the under-determined system in Eq. \eqref{E: basic tomography_stoch}, provided that the {\em sparsity} of the desired solution can be exploited. In other words, we are interested in solutions $\mathbf{x}$ with only a few, up to $k$, large entries. If all the other entries are small but non-zero, we refer to such vector as \emph{nearly} \emph{$k$-sparse}, and if they are exactly zero we call it \emph{exactly $k$-sparse}. Clearly, if vector $\mathbf{x}$ is exactly $k$-sparse, it is also nearly $k$-sparse. The appropriate definition that applies between \emph{nearly $k$-sparse} and \emph{$k$-sparse} will be clear from the context.

\begin{definition}\label{Def:k-identifiable}
A network is $k$-identifiable if for every {\em exactly $k$-sparse delay vector $\mathbf{x}$}, Eq. \eqref{E: basic tomography_stoch} is uniquely solvable.
\end{definition}

Compressed sensing \cite{jafarpour2009efficient,candes2006compressive,donoho2006compressed} has been proposed recently for network tomography \cite{coates2007compressed,xu2011compressive,wang2011sparse} as part of methods that vary significantly in their underlying assumptions and utility for practical networking scenarios. Authors in \cite{coates2007compressed} used compressed sensing to estimate link delays of the unobserved links on an end-to-end path when measured data is available on \emph{a subset of links}. Xu et. al. \cite{xu2011compressive} applied compressed sensing by performing a standard random walk over a \emph{sufficiently} connected graph to take measurements. However, this is at variance with typical network scenarios where the measurement matrix (i.e., routing matrix) is already given. Besides, most networks are not sufficiently connected \cite{faloutsos1999power,najiminaini2009analysis}. In this work we assume that the routing path between any pair of boundary nodes is predetermined without any constraint on the underlying network topology.

This work applies the concepts of {\em expander graphs} to the network tomography problem along with compressed sensing based link delay estimation. This is achieved by fundamentally relating the network routing matrix to a bipartite graph. If the bipartite graph is an {\em expander graph}, then one can use $l_1$ minimization to solve Eq. \eqref{E: basic tomography_stoch} for nearly $k$-sparse delay vector $\mathbf{x}$, that has polynomial complexity in $n$, independent of $k$ \cite{berinde2008sparse}.

\subsection{Contributions and Organization}\label{SS:contribution}

Our specific contributions are as follows. We first focus on 1-identifiable networks for the resulting intuition it generates. We relax the existing expansion result from $\epsilon< 1/6$ to $\epsilon \le 1/4$ (Lemma \ref{Tm:errorOfLP}) and extend our result to bipartite graphs that are a union of subgraphs, which are themselves expanders in Theorem \ref{Tm:extensionExpander}. Finally, in Theorems \ref{Tm:expander_tomo_regular_genK} and \ref{Tm:expander_tomo_genK} we show that for general k-identifiability, an inequality similar to 1-identifiable networks holds on average.

We conduct extensive simulation testing to show that the new results broaden the set of potential expanders (by up to 30\%) at the cost of accepting slightly larger error margin in reconstruction for the general $k$ case (Theorems \ref{Tm:expander_tomo_regular_genK} and \ref{Tm:expander_tomo_genK}). We derive the estimation error bound for $l_1$ minimization, and validate it via simulation results. From the results, it is evident that LP optimization under the new proposed conditions achieves better estimation accuracy as compared to techniques presented in the literature \cite{chen2010network,chen2007network}. This is due to the fact that our approach takes into account the inherent sparsity in the delay vector.

The rest of the paper is organized as follows: Section \ref{S:bipartite} relates the routing matrix of a
network to bipartite graphs. Section \ref{S:Exapdenr_Ident} establishes a connection
between link delay estimation and binary compressed sensing and
identifies conditions on the network routing matrix under which a given network is $k$-identifiable.
We evaluate our approach using simulations in
Section \ref{S:eval} and the paper concludes with  reflections on future work
in Section \ref{S:conl}. In the Appendix, we provide the proofs of the theorems.

\emph{Notations}: we use bold capitals (e.g., $\mathbf{R}$) to represent matrices and bold lowercase symbols
(e.g., $\mathbf{x}$) for vectors. The $i$-th entry of a vector $\mathbf{x}$ is denoted by
$x_i$. For the matrix $\mathbf{R}$, ${\cal N}(\mathbf{R})$ denotes its null space,
and superscript $^t$ denotes its transpose. Bold calligraphic capitalized symbols (e.g., $\mathRV{I}$) represents random variables. A set is denoted by a normal capital (e.g., $V$)
and a set of sets is presented by calligraphic capitalized symbol (e.g., $\mathcal{R}$) which is the set
of all end-to-end paths in the network.
$|\mathcal{R}|$ is the cardinality of the set. An empty set is denoted by $\emptyset$. $U \backslash A$ shows the set difference of $U$ and $A$. $deg(v)$ indicates degree of the node $v$ in a graph, defined as number of nodes it is connected to. For any set $S\subset\{1,2,\ldots,n\}$, $S^c$ represents the complement. Also, for any vector $\mathbf{x}\in \mathbb{R}^n$, vector $\mathbf{x}_S \, \in \, \mathbb{R}^n$ has entries defined as follows:
\begin{equation}
(x_S)_i=
\begin{cases}
x_i&\text{if  } i\in S\\
0&\text{ o.w. }
\end{cases}.
\end{equation}

If $\mathbf{x}\in\mathbb{R}^n$, the $l_p$-norm of $\mathbf{x}$, for $ p \, \ge \, 1$ is defined as $\parallel\mathbf{x}\parallel_p=\left(\sum_{i=1}^n |x_i|^p\right)^{1\over p}$.

\section{Routing Matrix and Bipartite Graph}\label{S:bipartite}
\begin{figure}
\centering
\begin{tabular}{cc}
\psfig{figure=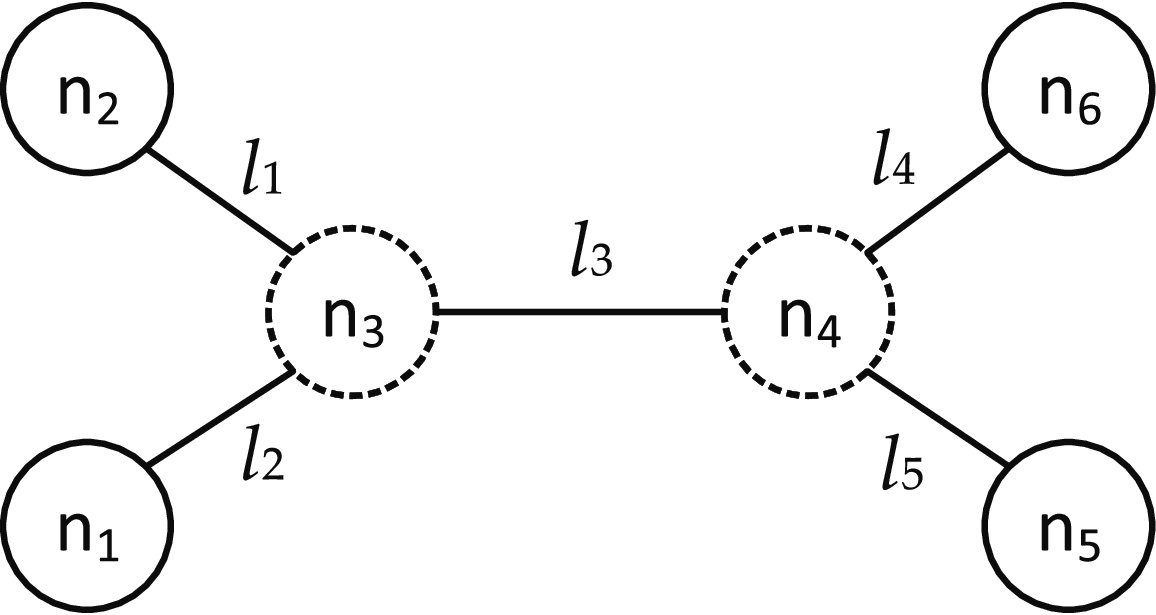,width=1.7in}
&
\psfig{figure=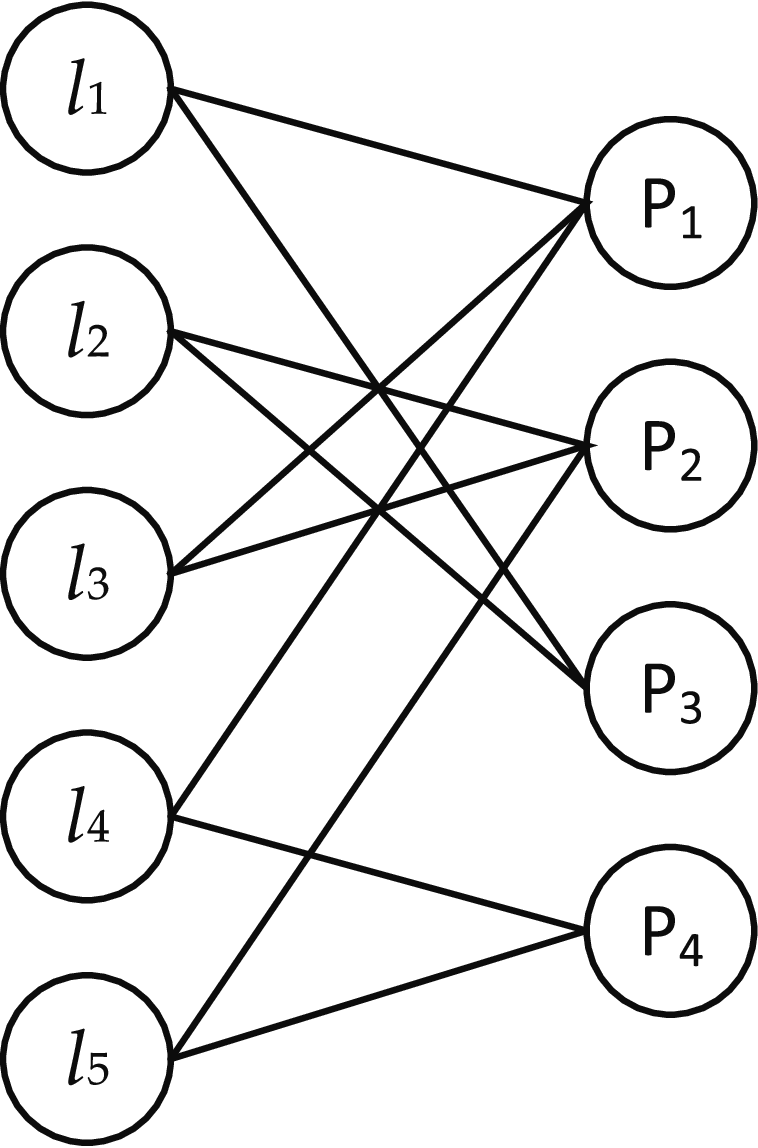,width=0.9in}\\
(\textbf{a})
&
(\textbf{b})
\end{tabular}
\caption{(a) A network with 4 boundary nodes, 2 intermediate nodes and 5 links, (b) Bipartite graph corresponding to given routing matrix in Eq. \eqref{E:P_exp}}
\label{F:2-2_example}
\end{figure}

As is customary, a network consisting of bidirectional links connecting transmitters, switches, and receivers can be modeled as an undirected graph $N(V,E)$, where $V$ ($E$) is the set of vertices (edges). Throughout the manuscript, boundary nodes are depicted as solid circles, while intermediate nodes are presented using dashed circles.  We use network depicted in Figure \ref{F:2-2_example}-(a) to illustrate the subsequent definitions.

A bipartite graph is one whose vertices can be divided into two disjoint sets, $X$ and $Y$, so that every edge connects a vertex in $X$ to one in $Y$ \cite{clark1991flg}. It is usually represented as a triple $G(X,Y,H)$, where $H\subset X\times Y$ is a set with paired elements from $X$ and $Y$. The vertex sets $X$ and $Y$ are the left and right sides of the graph, respectively. A bipartite graph $G(X,Y,H)$ can be represented by its {\em bi-adjacency} matrix $\mathbf{A}=[a_{ij}]$, where $a_{ij}=1$ if node $i \,\in\, Y $ is connected to node $j \, \in \,X$, and is zero otherwise. By definition, the elements in a row of a bi-adjacency matrix $\mathbf{A}$ correspond to $Y$ (right-hand side of the graph) and elements of columns correspond to $X$ (left-hand side of the graph). This convention is used throughout the paper.


Assume that a given network $N(V,E)$ has a total of $n$ links (i.e., $n=|E|$), and $\mathcal{R}$ is the (given) set of paths between the boundary nodes of the network and $r = |\mathcal{R}|$. Let $\mathbf{R}_{r\times n}$ denote the routing matrix corresponding to the set $\mathcal{R}$. For example, for the 1-identifiable network in Figure \ref{F:2-2_example}-(a), suppose the following routing matrix is given:
\begin{equation}
\mathbf{R} =
\begin{array}{cc}
   \begin{array}{c}
   \\
      P_1:n_2\rightsquigarrow n_6\\
      P_2:n_1\rightsquigarrow n_5\\
      P_3:n_1\rightsquigarrow n_2\\
      P_4:n_5\rightsquigarrow n_6
   \end{array}
   &
   \left[
   \begin{array}{ccccc}
      l_1&l_2&l_3&l_4&l_5 \\
      1&0&1&1&0\\
      0&1&1&0&1\\
      1&1&0&0&0\\
      0&0&0&1&1
   \end{array}
   \right],
\end{array}
\label{E:P_exp}
\end{equation}
which is equivalent to the following set of paths $\mathcal{R}$:
\begin{equation}
\mathcal{R} = \{l_1l_3l_4, l_2l_3l_5, l_1l_2, l_4l_5\}.
\end{equation}

$\mathbf{R}_{r\times n}$ can be viewed as a bi-adjacency matrix of a bipartite graph $G(X,Y,H)$, where $X=E$ (set of links in the network) and $Y=\mathcal{R}$ (set of given paths in the network). There exists a connection between a node in $X$ and a node in $Y$ if a path in $\mathcal{R}$ includes the corresponding link in $E$. Figure \ref{F:2-2_example}-(b) presents the bipartite graph for the network in Figure \ref{F:2-2_example}-(a) with the routing matrix $\mathbf{R}$ in Eq. \eqref{E:P_exp}.

Note that the above routing matrix, or its equivalent set of paths, is not a complete set of routes for the network in Figure \ref{F:2-2_example}-(a) (e.g., it does not include the path from $n_1$ to $n_6$, which is $l_2l_3l_4$). However, it is typically a fundamental premise in network tomography that the routing matrix is already chosen and may not be changed. Hence, we seek to investigate the following question: Assuming that the routing matrix is given, is it possible to estimate link delays?

\section{Expander Graphs and Network Identifiability}\label{S:Exapdenr_Ident}

In recent years, a new approach--{\em Compressed Sensing}--for estimating an $n$-dimensional (signal) vector $\mathbf{x}$ from a lower-dimensional representation has attracted much attention \cite{donoho2006compressed,candes2006robust,candes2006compressive}. For any signal $\mathbf{x} \; \in \; {\mathbb R}^n$, the reduced dimension representation is equal to $\mathbf{y} = \mathbf{Ax}$, where $m\times n$ matrix $\mathbf{A}$ ($m \ll n$) is referred to as the \emph{measurement matrix}. The main challenge in traditional compressed sensing is to construct $\mathbf{A}$ with the following desirable (and conflicting) properties: (a) achieve maximum possible compression ($m/n$ small) and yet allow (b) an accurate reconstruction of $\mathbf{x}$ from $ \mathbf{y}$ when $\mathbf{x}$ is known to be sparse using (c) a fast decoding algorithm \cite{wakin2008introduction,gilbert2007one,candes2006near,xu2007efficient}.

As discussed above, the routing matrix of a network is the measurement matrix for delay tomography application, and in most scenarios it is predetermined. The main issue, therefore, is to determine whether it is an \emph{appropriate} measurement matrix for compressed sensing, i.e., if it satisfies objective (b) above. In the simulation section, we show that the existing conditions for the measurement matrix do \emph{not} apply to most of the routing matrices. Motivated by this observation, we aim to revisit these conditions and modify them so that they become more suitable to the network tomography problem. 
\subsection{Expander Graphs}
\begin{definition}\label{Def:Expander}
A left $d$-regular bipartite graph $G(X,Y,H)$; i.e., $deg(v)=d\;\forall\,v \in X$,  is a $(\phi,d,\epsilon)$-expander if for any $\Phi\subset X$ with $|\Phi|\le \phi$, the following condition holds:
\begin{equation}
|N(\Phi)|\ge \; (1-\epsilon)d \, |\Phi|,
\label{E:expansionProperty}
\end{equation}
where $N(\Phi)$ is a set of neighbors of $\Phi$. Neighbors of $\Phi$ are nodes which are connected to at least one of the nodes in $\Phi$. $\phi$ and $\epsilon$ are the ``expansion factor'' and the ``error parameter,'' respectively.
\end{definition}

In an expander graph, the degree of connectivity for a set of nodes (with cardinality of up to $\phi$) on the left-hand side ($X$) expands by the factor $(1-\epsilon)d$ on the right-hand side ($Y$) \cite{sarnak2004expander}. Expander graphs are well-studied; in a key result, Berinde and Indyk in \cite{berinde2008combining,berinde2008sparse} show that the bi-adjacency matrix of a $(2\phi,d,\epsilon)$-expander graph can be used as the measurement matrix for a $\phi$-sparse signal, for $\epsilon < {1\over 6}$. The parameter $\epsilon$ in an expander graph is a design variable that is related to recovery error. The existing results for a bipartite graph to be an expander require $\epsilon < 1/6$, which, as we will show, does not apply to most networks. For the network tomography problem, the measurement matrix is usually pre-determined, and we need to enlarge the bound on $\epsilon$ so as to increase the likelihood that it leads to an identifiable network.

The bipartite graph in Figure \ref{F:2-2_example}-(b) corresponds to the 1-identifiable network in Figure \ref{F:2-2_example}-(a) with the routing matrix in Eq. \eqref{E:P_exp}. It is easy to see that this bipartite graph is an expander for $\epsilon = 1/4$. Motivated by this example, we relax the existing result for $\epsilon < 1/6$ to $\epsilon \le 1/4$. In the simulation results (Section \ref{SS:Net_expansion_Sim}), we show that this relaxation increases the number of $k$-identifiable networks that satisfy the expansion property. For networks that satisfy the expansion property, LP optimization can be used to solve the tomography problem. The analytical results are first derived for 1-identifiable networks and then generalized to $k > 1 $-identifiability.
\subsection{1-identifiability}\label{S:networkIdent}
Consider a network $N(V,E)$ with a routing matrix $\mathbf{R}$ that satisfies the expansion property of a $(2, d, \epsilon)$-expander graph. Now, let us consider two links that constitute two nodes on the left-hand side of bipartite graph $G(E,\mathcal{R},H)$; i.e., $|\Phi|=2$. As indicated by the definition of expander graphs in Eq. \eqref{E:expansionProperty}, the number of nodes on the right-hand side connected to these two nodes is $2d(1-\epsilon)$. If the two left nodes are connected to exactly the same nodes on the right-hand side, distinguishing between them is impossible,  and the bi-adjacency matrix $\mathbf{R}$ is rank deficient. Identifying the correct congested link requires that the number of paths passing through these two links be greater than $d$. Thus, $2d(1-\epsilon)\ge d+1$ for any $d$, which for $d=2$ implies $\epsilon \le \frac{1}{4}$. We exclude case of $d = 1$ because each left 1-regular bipartite graph with $N(\Phi) \ge 2$ is an expander graph.
Following the above, Lemma \ref{Tm:relax_epsilon} provides an upper bound on the error of recovering $\mathbf{x}$ from its lower-dimensional projection $\mathbf{Ax}$ when $\mathbf{A}$ is a bi-adjacency matrix of a $(2,d,\epsilon)$-expander graph and $\epsilon\le 1/4$.

\begin{lemma}
Let $\mathbf{A}$ be a bi-adjacency matrix of a $(2,d,\epsilon)$-expander graph with $\epsilon\le 1/4$. Consider any two vectors, $\mathbf{x}$ and $\mathbf{x}^\prime$, with the same projection under the measurement matrix $\mathbf{A}$, i.e., $\mathbf{Ax}=\mathbf{A}\mathbf{x}^\prime$. Assume that $\mathbf{x}$ is 1-sparse and without loss of generality, $\parallel \mathbf{x}^\prime\parallel_1\le \parallel \mathbf{x}\parallel_1$. Let $S$ be the singleton set of the largest (in magnitude) element of $\mathbf{x}$. Then,
\begin{equation}
\parallel \mathbf{x}^\prime-\mathbf{x}\parallel_1 \le f(\epsilon)\parallel \mathbf{x}_{S^c}\parallel_1,
\label{E:xp_fe}
\end{equation}
\label{Tm:errorOfLP}
where
\begin{equation}
f(\epsilon)={{2(1+2\epsilon)}\over {1-2\epsilon}},\;\; \epsilon\le {1 \over 4}.
\label{E:fe}
\end{equation}
\label{Tm:relax_epsilon}
\end{lemma}

\emph{Proof}: see Appendix\footnote{In \cite{berinde2008combining,berinde2008sparse}, this function is  derived as $f(\epsilon)=\frac{2(1-2\epsilon)}{1-6\epsilon}$; thus, it requires $\epsilon < \frac{1}{6}$.}.

The above lemma suggests that under some conditions, the link delay in a network may be correctly estimated from the measurements represented by Eq. \eqref{E: basic tomography_stoch}. If $\mathbf{x}^\prime$ represents the estimate (via a suitable algorithm) of the true (unknown) delays $\mathbf{x}$, the lemma suggests that the estimation error is appropriately bounded, whenever the measurement matrix $\mathbf{A}$ is a bi-adjacency matrix of an expander graph.

The following theorem relates the problem of delay estimation in a network $N(V,E)$ to results on expander graphs with $\epsilon\le 1/4$ and shows that Eq. \eqref{E: basic tomography_stoch} can be solved for $\mathbf{x}$ using LP optimization.

\begin{theorem}
Let $N(V,E)$ be a network with a set of paths $\mathcal{R}$ and a corresponding routing matrix $\mathbf{R}_{r\times n}$. Suppose that $G(E,\mathcal{R},H)$ is a bipartite graph with bi-adjacency matrix $\mathbf{R}$. Assume that $\mathbf{x}$ is the true (unknown) 1-sparse delay vector of $N(V,E)$ and $\mathbf{y}=\mathbf{Rx}$ is the (given) end-to-end delay measurement. Let $\mathbf{x}^\prime$ be a solution to the following LP optimization:
\begin{eqnarray}\label{E:LP}
&&\min \parallel \mathbf{x}^\prime\parallel_1 \\ \nonumber
&&\text{s.t.} \\ \nonumber
&&\mathbf{R}\mathbf{x}^\prime=\mathbf{y}.
\end{eqnarray}

Then,
\begin{equation}\label{E:upper_bound}
\parallel \mathbf{x}-\mathbf{x}^\prime\parallel_1\le f(\epsilon)\parallel \mathbf{x}_{S^c}\parallel_1,
\end{equation}
if $G$ is a $(2,d,\epsilon)$-expander with $\epsilon\le {1\over 4}$.
\label{Tm:expander_tomo}
\end{theorem}
\emph{Proof}: see Appendix.

Note that if the true delay vector $\mathbf{x}$ is exactly 1-sparse (which rarely occurs), it implies that $\parallel \mathbf{x}_{S^c}\parallel_1=0$, which means that $\mathbf{x}^\prime=\mathbf{x}$; i.e., $l_1$-norm minimization in Eq. \eqref{E:LP} can recover $\mathbf{x}$ with zero estimation error. In other words,  if the delay of all links in the network is zero except for one, the delay of that link can be exactly recovered from the end-to-end delay measurement. However, if the true delay vector contains links with small but nonzero delays (the more likely scenario), the above theorem yields an upper bound on the estimation error.

\begin{figure}
\centering
\begin{tabular}{cc}
\psfig{figure=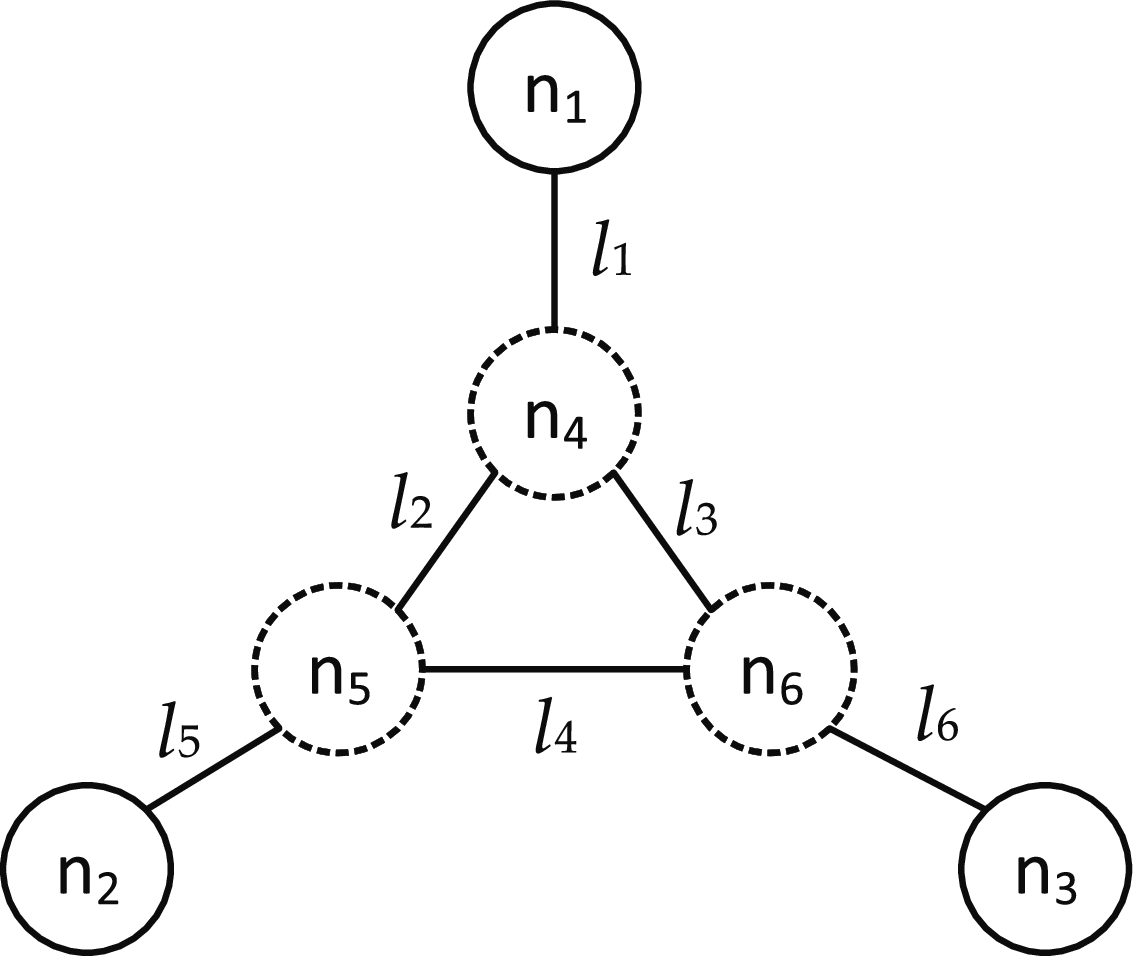,width=1.7in}
&
\psfig{figure=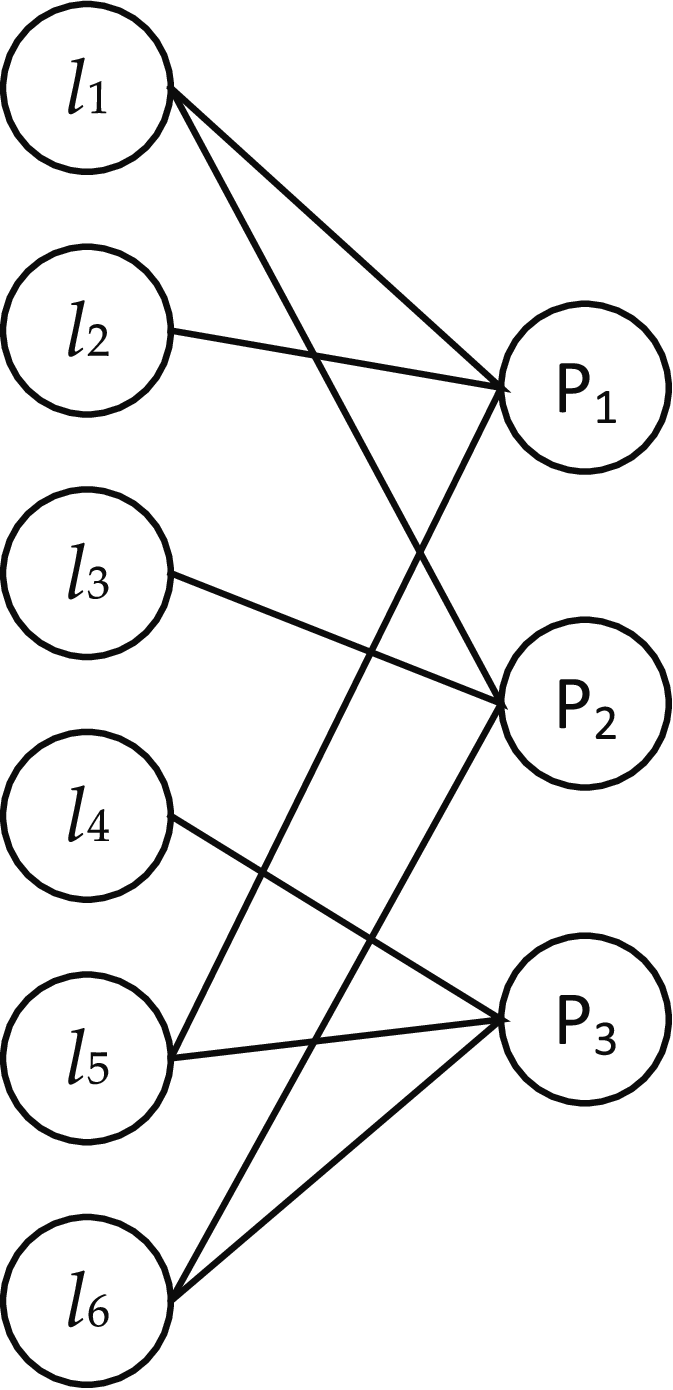,width=0.9in}
\\
\textbf{(a)}&\textbf{(b)}
\end{tabular}
\caption{An example of 1-identifiable network whose corresponding bipartite graph
is not left regular. (a) Network topology (b) Bipartite graph corresponding to the routing matrix $\mathbf{R}$ in Eq. \eqref{E:nonregular_exp}}
\label{F:NotMulticast}
\end{figure}

\emph{Relaxing $d$-regularity Condition}: our result requires expander graphs that are left $d$-regular. However, there exists networks which are $1$-identifiable, but their corresponding bipartite graphs are not left $d$-regular, suggesting that the above result is sufficient but not necessary. An example of such a network is depicted in Figure \ref{F:NotMulticast}-(a) with the following routing matrix:
\begin{equation}
\mathbf{R} =\!\!\!\!
\begin{array}{cc}
   \begin{array}{c}
   \\
      P_1:n_1\rightsquigarrow n_2\\
      P_2:n_1\rightsquigarrow n_3\\
      P_3:n_2\rightsquigarrow n_3\\
   \end{array}\!\!\!\!
   &
   \left[
   \begin{array}{cccccc}
   l_1&l_2&l_3&l_4&l_5&l_6\\
      1&1&0&0&1&0\\
      1&0&1&0&0&1\\
      0&0&0&1&1&1\\
   \end{array}
   \right].
\end{array}
\label{E:nonregular_exp}
\end{equation}

\begin{figure}
\centering
\begin{tabular}{cc}
\psfig{figure=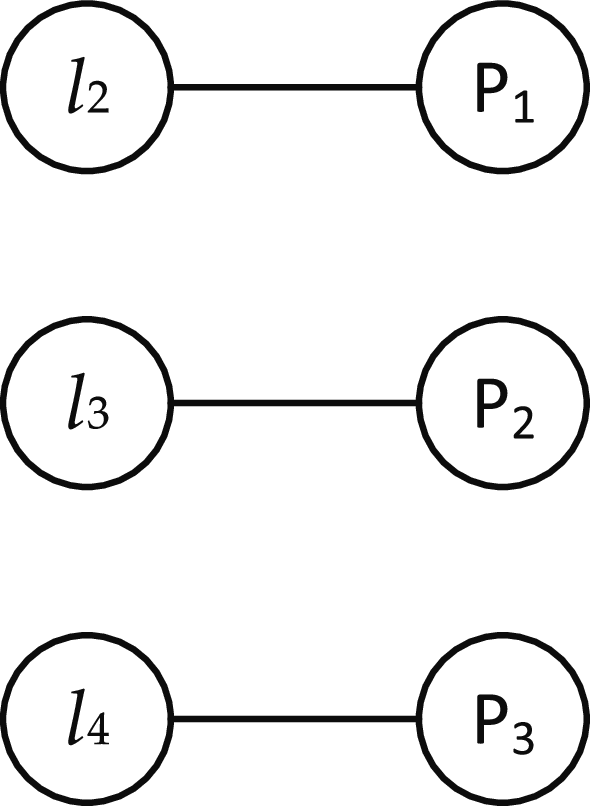,width=0.9in}
&
\psfig{figure=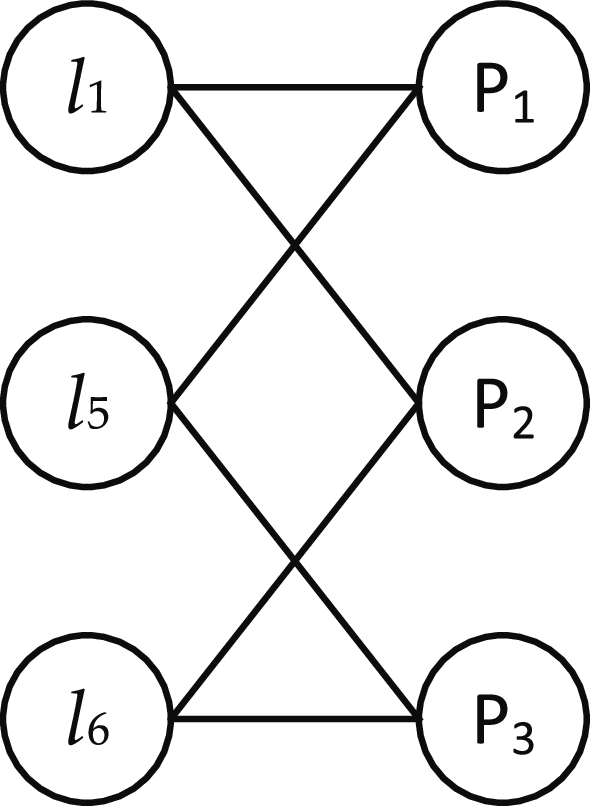,width=0.9in}
\\
\textbf{(a)}&\textbf{(b)}
\end{tabular}
\caption{Two subgraphs of the bipartite graph in Figure \ref{F:NotMulticast}-(b) which are left regular.}
\label{F:bipartite6x3}
\end{figure}

The above routing matrix is a bi-adjacency matrix of the bipartite graph presented in Figure \ref{F:NotMulticast}-(b). This bipartite graph is not left regular because the degree of a node in the left set is either 1 or 2; thus, it cannot be an expander. However, Figures \ref{F:bipartite6x3}-(a) and (b), respectively, represent subgraphs of $G$ with regular left degree 1 and 2; these subgraphs are expander graphs. The above observation suggests that the result in Theorem \ref{Tm:expander_tomo} may be extended to networks whose corresponding bipartite graph is not regular (and is therefore not an expander) but can be partitioned into disjoint union of subgraphs that are themselves expander graphs.

\begin{theorem}
Let $N(V,E)$ be a network with routing matrix $\mathbf{R}_{r\times n}$. Let $G(X,Y,H)$ be a bipartite graph with bi-adjacency matrix $\mathbf{R}$. Suppose that $G_i(X_i,Y,H_i)$, $i=1,2,\ldots,M$, are $d_i$-regular bipartite subgraphs of $G$ such that
 \begin{itemize}
 \item $X=\cup X_i$ and $X_i\cap X_j = \emptyset$ for $i\not= j$
 \item $H=\cup H_i$
 \item $d_i\not= d_j$ for $i\not= j$
 \end{itemize}

Then, $N(V,E)$ is $1$-identifiable if $G_i$ is a $(2,d_i,\epsilon)$-expander
graph with $\epsilon\le {1 \over 4}$, $i=1,2,\ldots, M$. Further, the link delay is
the solution to the LP optimization in Eq. \eqref{E:LP}.
\label{Tm:extensionExpander}
\end{theorem}

\emph{Proof}: see Appendix.

For future reference, we refer to the conditions in Theorem \ref{Tm:extensionExpander} as \emph{1-identifiability expansion conditions}. If the network $N(V,E)$ satisfies these conditions, we refer to it as the \emph{$1$-identifiable expander network}. The conditions in Theorem \ref{Tm:extensionExpander} imply the following for any link pair $l_i$ and $l_j$:

\begin{itemize}
\item They belong to different $G_i$'s and hence have different degrees $deg(l_i)\not=deg(l_j)$.
\item They belong to the same subgraph $G_i$, i.e., $deg(l_i)=deg(l_j)=d_i$. In that case, because $G_i$ is a bipartite graph, they satisfy  the expansion property in Eq. \eqref{E:expansionProperty}.
\end{itemize}

We state this observation formally in the following corollary.

\begin{corollary}
Let $N(V,E)$ be a 1-identifiable expander network with the routing matrix $\mathbf{R}_{r\times n}$ and a set of paths $\mathcal{R}$. Let $G(E,\mathcal{R},H)$ be its corresponding bipartite graph with the bi-adjacency matrix $\mathbf{R}$. Then, one and only one of the following statements is true for any two links $l_i$ and $l_j$ in $E$, $i\not=j$:
\begin{itemize}
\item $deg(l_i)>deg(l_j)$
\item $deg(l_i)<deg(l_j)$
\item $deg(l_i)=deg(l_j)=d$ and $2d-4deg(l_i,l_j)\ge0$
\end{itemize}
\label{Tm:identifiabilityConditions}
\end{corollary}
where $deg(l_i,l_j)$ is defined as the number of nodes connected to both $l_i$ and $l_j$ in the bipartite graph $G$. Equivalently, $deg(l_i,l_j)$ denotes number of paths going through both $l_i$ and $l_j$.

\subsection{k-identifiability}

For $k$-identifiable networks, the performance of $l_1$ optimization in terms of estimation error, depends on the location and number of the congested links. It thus does not seem feasible to provide strict {\em deterministic} counterparts to Theorems 1, 2 for $k$-identifiability in general. Instead, we prove - using a probabilistic approach - that the {\em average estimation error over the distribution of congested links (up to $k$) within the network} is upper bounded, and provide a bound in terms of the delays of the (other) uncongested links.

To do so, we follow a procedure similar to $k=1$  in Lemma \ref{Tm:errorOfLP}, Theorems \ref{Tm:expander_tomo}, and \ref{Tm:extensionExpander}. We show that when $\mathbf{R}$  is a bi-adjacency matrix of a $(2k, d, \epsilon)$-expander graph with $\epsilon\le 1/4$, the LP optimization in Eq. \eqref{E:LP} can be used for recovery of delay vector $\mathbf{x}$. Similar to Eq. \eqref{E:upper_bound}, we proof that the expected estimation error is bounded by average delay of uncongested links in the network. Here, we assume that when a link is uncongested, it has a constant small delay value. This is a fair assumption because the delay of the uncongested links is primarily dominated by the propagation delay \cite{bovy2002analysis}, which is constant. Furthermore, for the sake of simplicity, we also assume that when a link is congested it has a constant (unknown) large delay value.

A $k$-sparse delay vector $\mathbf{x}$, contains up to $k$ large entries and the rest are close to zero. We define the random set $\mathRV{S}$ as the set of indexes of the large entries in $\mathbf{x}$. Thus, $\mathRV{S}\subset\{1,2,\ldots,n\}$, $|\mathRV{S}|\le k$ and $\mathRV{S}^c=\{1,2,\ldots,n\}\backslash \mathRV{S}$. In network $N(V,E)$, random set $\mathRV{S}$ represents the congested links. A realization of $S$ can be denoted as $S=\{i_1,i_2,...,i_j\}, j\le k$ where $x_{i_1},x_{i_2},...,x_{i_j}$ are the large values in vector $\mathbf{x}$ and the rest are close to zero. Accordingly, links $l_{i_1}, l_{i_2},\ldots, l_{i_j}$ are congested. As we will see in the following theorem, the performance of using the optimization in Eq. \eqref{E:LP} does not depend on the values of $x_{i_1},x_{i_2},...,x_{i_j}$.

\begin{theorem}
Let $N(V,E)$ be a network with a set of paths $\mathcal{R}$ and a corresponding routing matrix $\mathbf{R}_{r\times n}$. Suppose that $G(E,\mathcal{R},H)$ is a $(2k,d,\epsilon)$-expander graph with bi-adjacency matrix $\mathbf{R}$ and $\epsilon\le \frac{1}{4}$.  Assume that $\mathbf{x}$ is the unknown $k$-sparse delay vector of $N(V,E)$, $\mathbf{y}=\mathbf{Rx}$ is the given end-to-end delay measurement, and $\mathbf{x}^\prime$ is the solution to the LP optimization in Eq. \eqref{E:LP}. Then, the expected estimation error is bounded as follows:
\begin{equation}\label{E:upper_bound_genK}
\mathbb{E}_{\mathRV{S}}\left[\parallel \mathbf{x}_{\mathRV{S}}-\mathbf{x}^\prime_{\mathRV{S}}\parallel_1+\parallel \mathbf{x}_{\mathRV{S}^c}-\mathbf{x}^\prime_{\mathRV{S}^c}\parallel_1\right]\le f(\epsilon)\mathbb{E}_{\mathRV{S}}[\parallel \mathbf{x}_{{\mathRV{S}}^c}\parallel_1],
\end{equation}
where $\mathRV{S}$ ($|\mathRV{S}|\le k$) is the set of congested links inside $N(V,E)$ and the expectation is with respect to the distribution of $\mathRV{S}$.
\label{Tm:expander_tomo_regular_genK}
\end{theorem}
\emph{Proof}: see Appendix.

The probability distribution of $\mathRV{S}$ can be very complex in general. One can assume the prior distribution for congestion events and independency among the links' congestion events. To monitor the network, however, there is no need to know the distribution of $\mathRV{S}$ in advance. If a network $N(V,E)$ satisfies the $k$-identifiability condition (given in Definition \ref{Def:k-identifExpander}), the network operator or the ISP owner can use the LP optimization in Eq. \eqref{E:LP} to recover the delay in the network using end-to-end monitoring.

We define a $k$-identifiable expander network as follows:

\begin{definition}\label{Def:k-identifExpander}
A $k$-identifiable expander network $N(V,E)$ is a network whose routing matrix $\mathbf{R}_{r\times n}$ is the bi-adjacency matrix of a bipartite graph $G(X,Y,H)$ consisting of $d_i$-regular subgraphs $G_i(X_i,Y,H_i)$ with the following properties
\begin{itemize}
\item $X=\cup X_i$ and $X_i\cap X_j = \emptyset$ for $i\not= j$
\item $H=\cup H_i$
\item $d_i\not= d_j$ for $i\not= j$
\item $G_i(X_i,Y,H_i)$ is a $(2k,d_i,\epsilon)$-expander with $\epsilon\le {1\over 4}$
\end{itemize}
\end{definition}

Theorem \ref{Tm:expander_tomo_regular_genK} holds for networks whose corresponding bipartite graphs are left regular. In  following theorem we relax this condition. We show that for any $k$-identifiable expander network $N(V,E)$ with a given routing matrix $\mathbf{R}$, its $k$-sparse delay vector $\mathbf{x}$ can be recovered from end-to-end measurement $\mathbf{y}=\mathbf{Rx}$  with LP optimization in Eq. \eqref{E:LP}.

\begin{theorem}
Let $N(V,E)$ be a $k$-identifiable expander network with a set of paths $\mathcal{R}$ and a corresponding routing matrix $\mathbf{R}_{r\times n}$. Then the $k$-sparse delay vector of $N(V,E)$ is the solution to the LP optimization in Eq. \eqref{E:LP}.
\label{Tm:expander_tomo_genK}
\end{theorem}
\emph{Proof}: see Appendix.

\section{Evaluation Results}\label{S:eval}
In Section \ref{S:Exapdenr_Ident}, Theorem \ref{Tm:expander_tomo_genK}, we showed that if the routing matrix of a network is the bi-adjacency matrix of the union of disjoint expander graphs ($k$-identifiable expander as defined in Definition \ref{Def:k-identifExpander}), that network is $k$-identifiable. Moreover, we can estimate internal link delay using an LP optimizer in Eq. \eqref{E:LP}. However, a legitimate 'big-picture' question arises: How many networks actually satisfy the conditions of Theorem \ref{Tm:expander_tomo_genK}; i.e., how many are $k$-identifiable expanders? In this section, we generate random Internet-type networks to study this question. Our simulation results show that our relaxation increases number of networks which satisfy expansion property by almost 30\%.

For those networks that are $k$-identifiable expander--i.e., their routing matrix satisfies the condition in Definition \ref{Def:k-identifExpander}--we determine the average normalized estimation error when there is $k$ congested link in the network and show that the average normalized estimation error remains within an acceptable range. Next, we compare $l_1$ optimization with new conditions with a recently developed delay tomography algorithm and show that the LP optimization under the new proposed conditions yields a lower estimation error.

Finally, some network monitoring applications are required to locate congested links in a network. We show that if the network is $k$-identifiable, the LP optimization algorithm can be employed as a congestion detection tool that has a performance close to optimum detector.

\subsection{Generation of Networks with Random Topology}\label{S:GeneratedNet}
\begin{figure*}
\begin{tabular}{cc}
\psfig{figure=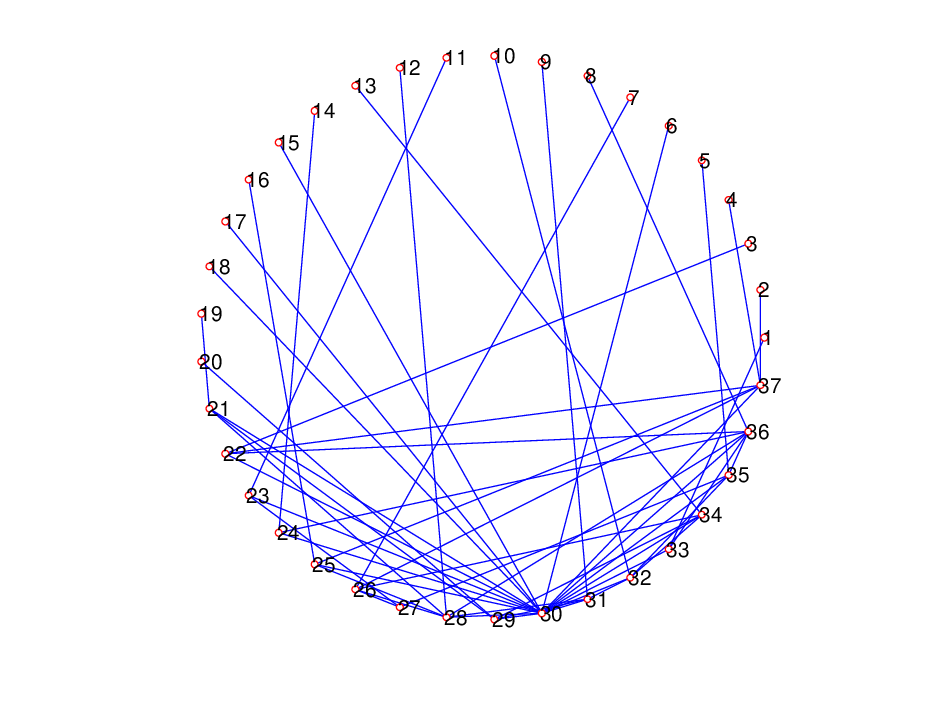,width=3.0in}&
\psfig{figure=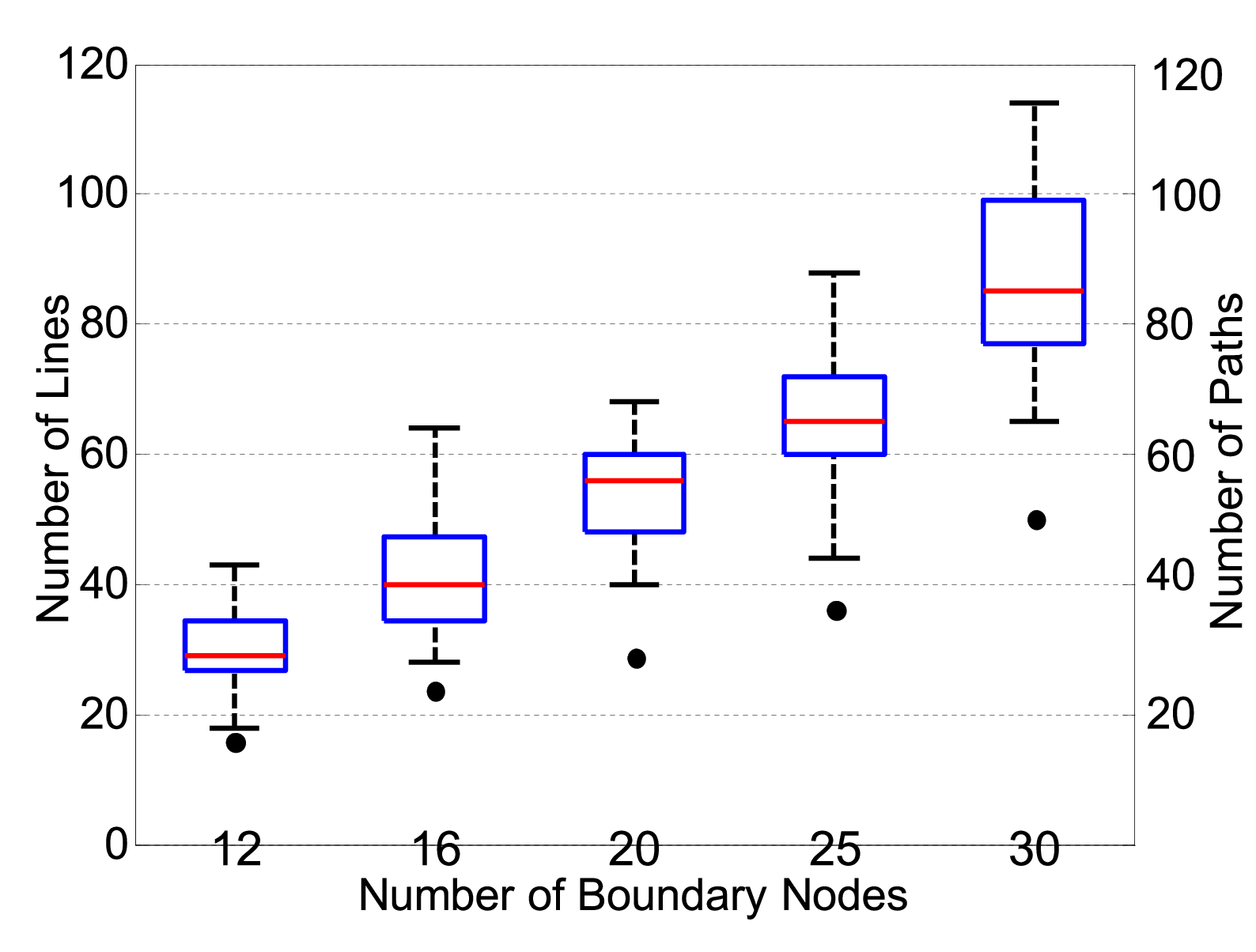,width=3.0in}\\
(\textbf{a})&(\textbf{b})
\end{tabular}
\caption{(a) Output of Inet after our modification in MATLAB with 20 boundary nodes. Nodes with degree-1 are injection nodes. (b) Statistics of number of links within a network with fix number of boundary nodes. For each number of boundary nodes, the central mark is the median, the edges of the box are the 25th and 75th percentiles, and upper/lower bars represent the extreme values observed. The filled circle shows the average number of paths used for delay recovery.}
\label{F:InetGraphs}
\end{figure*}

We use Inet version 3.0 \cite{winick2002inet}--an Internet topology generator software at AS (Autonomous System) level-- to generate random graphs with the given power law factor and a fixed number of boundary nodes. Boundary nodes are nodes with degree one which act as injection points for probes in network tomography problem. We create networks containing 5000 nodes with 12, 16, 20, 25 and 30 boundary nodes, respectively. The output of Inet, which contains the set of neighbors of each node in the generated graph, is fed to matgraph toolbox in MATLAB \cite{sheinerman2007matgraph} for modification. We first create a routing matrix containing the shortest paths between any boundary node pairs in the network. Then we delete all nodes and links that do not contribute to any of the above paths, since  if a link is not covered by any end-to-end path, it is not identifiable. The remaining networks constitute our random set. In Figure \ref{F:InetGraphs}-(a), an example of random network is depicted.

It is worth pointing out that in Internet topology-based networks, a relationship exists between the number of boundary nodes (number of nodes with degree one) and the number of links in a network (often referred to as the network size) \cite{chen2003tomography}. The literature shows that the degree of nodes in Internet topology has a power law distribution. The majority of Internet topology generator softwares, including Inet 3.0, consider the number of boundary nodes as an input argument \cite{winick2002inet}. In Figure \ref{F:InetGraphs}-(b) Statistics of number of links within a network with fix number of boundary nodes is depicted. Clearly, size of the networks (number of links in the networks) grows with the number of boundary nodes. Moreover, In Figure \ref{F:InetGraphs}-(b) presents the average number of paths in the networks that are available for delay recovery. As expected, number of paths in the network is less than number of links.

\subsection{Networks and Expansion Property}\label{SS:Net_expansion_Sim}

For the routing matrices of these random networks, we first examine how many of them satisfy the $k$-identifiability  expansion conditions in Definition \ref{Def:k-identifExpander}. For the network with a fixed number of boundary nodes, fifty different topologies are created. Table \ref{T:PercOfExpander_<1/6&<1/4} shows the percentage of networks that satisfy the $k$-identifiability property for $k=1,2,3$.

To show the impact of our relaxation of $\epsilon$ in Lemma \ref{Tm:errorOfLP} and Theorem \ref{Tm:expander_tomo_genK}, in Table \ref{T:PercOfExpander_<1/6&<1/4}, we also provide the percentage of networks that are $k$-identifiable, using $\epsilon < 1/6$. As one can see, by moving the bound on $\epsilon$ from $1/6$ to $1/4$, the number of networks satisfying the expansion property increases by almost 30\%. In other words, 30\% of $k$-identifiable expander networks are within $1/6\le \epsilon\le 1/4$.

\begin{table}
\centering
\caption{For networks with a fixed number of boundary nodes, how many are $k$-identifiable expanders for $k=1,2,3$ with $\epsilon\le {1 \over 4}$ and $\epsilon < {1\over 6}$}
\begin{tabular}{cc|c|c|c|c|c|}
\cline{3-7}
& & 12 & 16 & 20 & 25 & 30\\
\hline
\multicolumn{1}{|c|}{\multirow{2}{*}{$k=1$}}
& \multicolumn{1}{|c|}{$\epsilon\le {1\over 4}$} & 72\% & 74\% & 72\% & 70\%& 72\%\\ \cline{2-7}
\multicolumn{1}{|c|}{}
& \multicolumn{1}{|c|}{$\epsilon< {1\over 6}$} & 40\% & 32\% & 30\% & 36\%& 32\%\\ \cline{1-7}\hline\hline
\multicolumn{1}{|c|}{\multirow{2}{*}{$k=2$}}
& \multicolumn{1}{|c|}{$\epsilon\le {1\over 4}$} & 56\% & 50\% & 50\% & 54\%& 52\%\\ \cline{2-7}
\multicolumn{1}{|c|}{}
& \multicolumn{1}{|c|}{$\epsilon< {1\over 6}$} & 20\% & 24\% & 22\% & 22\% & 26\%\\ \cline{1-7}\hline\hline
\multicolumn{1}{|c|}{\multirow{2}{*}{$k=3$}}
& \multicolumn{1}{|c|}{$\epsilon\le {1\over 4}$} & 56\% & 50\% & 46\% & 52\%& 52\%\\ \cline{2-7}
\multicolumn{1}{|c|}{}
& \multicolumn{1}{|c|}{$\epsilon< {1\over 6}$} & 0\% & 0\% & 22\%  & 20\%& 24\%\\ \cline{1-7}
\end{tabular}
\label{T:PercOfExpander_<1/6&<1/4}
\end{table}

\subsection{Delay Estimation: Simulation Experiments}\label{S:delayEstimation}
Theorem \ref{Tm:expander_tomo_genK} says that if the routing matrix of a network satisfies $k$-identifiable expander conditions, then link delays in the network can be estimated using Eq. \eqref{E:LP} with bounded estimation error. To examine the accuracy of the proposed delay estimation method, for each network created in Section \ref{S:GeneratedNet}, we calculate the average normalized estimation error for all links as follows. When a link is uncongested, it experiences fixed delay value sampled from exponentially distributed delays with average $\mu$, i.e.,
\begin{equation}\label{E:f_l}
f_l(t)={1\over \mu}exp(-{t\over \mu})\;\;\;\forall l\in E,
\end{equation}
where $f_l(t)$ is the delay for link $l$ and $\mu \; \in \; [0, 1]$ to denote that these links do not undergo congestion. To model congestion events, $k$ reference links are randomly selected and assigned a delay of $10$ ms to denote congestion.

We exploit the LP optimization in Eq. \eqref{E:LP} to estimate link delays for $k$-identifiable expander networks. For a network, the normalized estimation error for each congested link inside the network is calculated as follows:
\begin{equation}
norm. \; err={{\parallel\mathbf{x}-\hat{\mathbf{x}}\parallel_1}\over {\parallel\mathbf{x}\parallel_1}},
\end{equation}
where $\mathbf{x}$ and $\hat{\mathbf{x}}$ are the true and estimated delay vectors respectively.

\begin{figure*}
\centering
\begin{tabular}{cc}
\psfig{figure=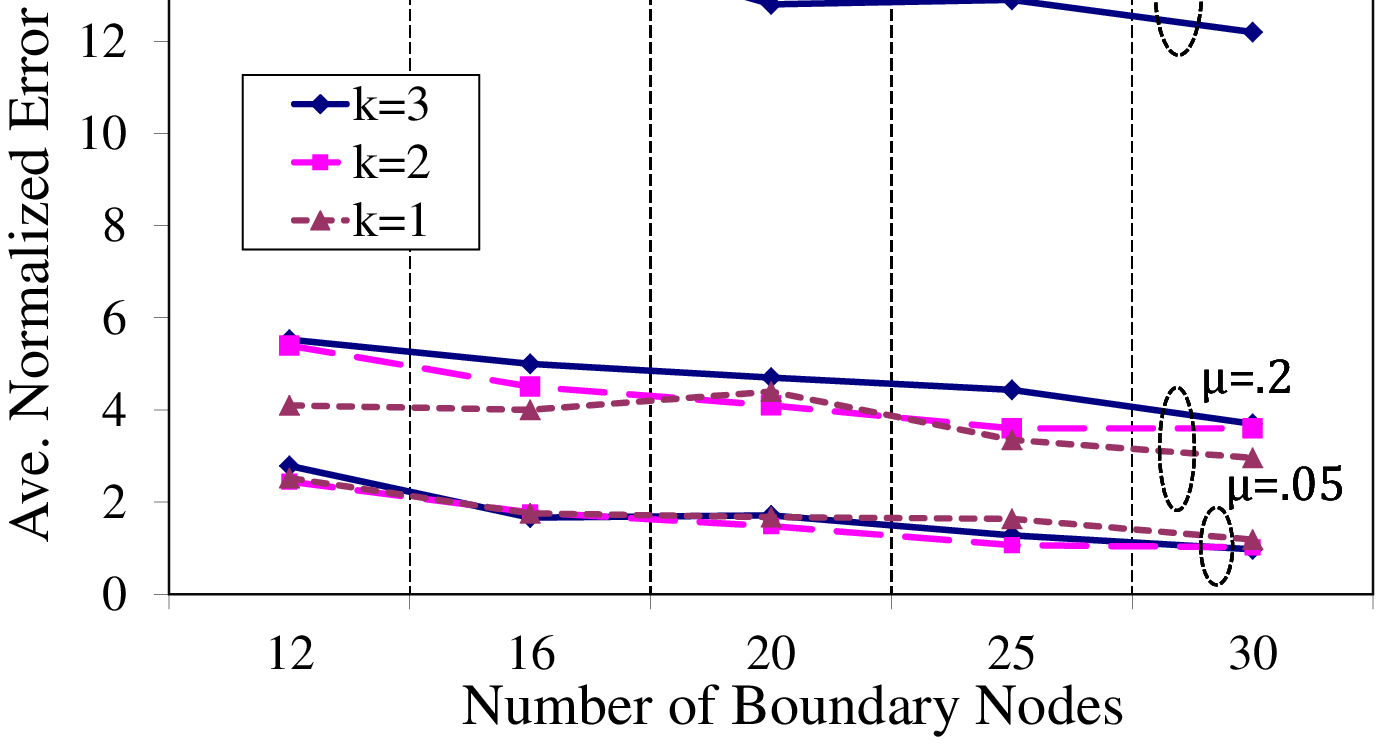,width=3.15in,rheight=1.3in}& \psfig{figure=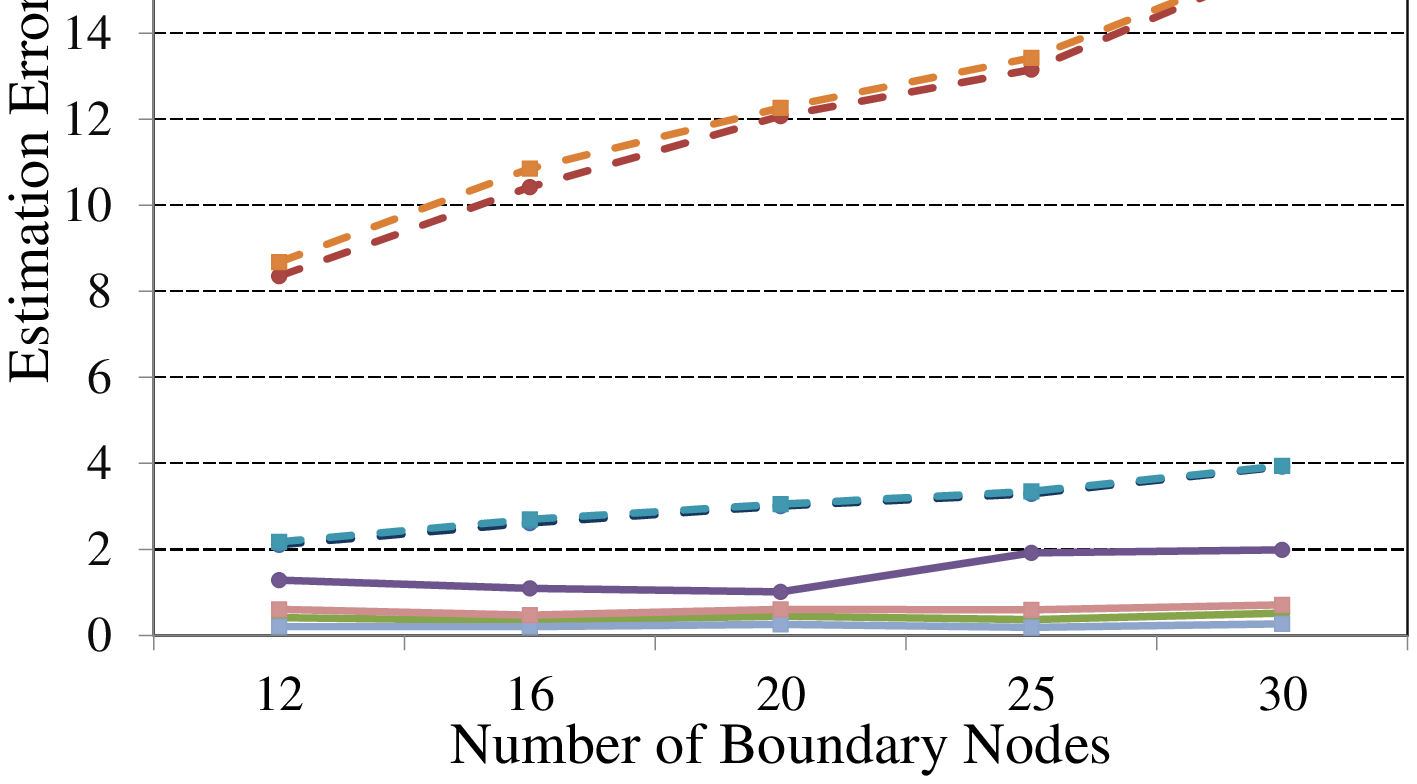,width=3.0in,rheight=1.3in}\\
(\textbf{a})&(\textbf{b})
\end{tabular}
\caption{(a) Average normalized estimation error in networks satisfying conditions given in Definition \ref{Def:k-identifExpander} when there are $k$ deficient link within the network for different values of $\mu$ in Eq. \eqref{E:f_l}. (b) Comparing the simulation estimation error and the derived theoretical error upper bound.}
\label{F:error}
\end{figure*}

Figure \ref{F:error}-(a) presents the average normalized estimation error when there are $k$ congested links inside the network for $k=1,2,3$ and LP optimization Eq. \eqref{E:LP} is used to estimate the delay. As expected, the average normalized estimation error for different $\mu$ (vector $ {\bf x}$ is nearly $k$-sparse) mimics the expected trend from Eq. \eqref{E:upper_bound_genK}; i.e., as $\mu$ decreases (the average delay of uncongested links goes to zero), so does the recovery error of the LP algorithm. This phenomena also can be seen in Figure \ref{F:error}-(b) which presents the estimation error upper bound. From Eq. \eqref{E:upper_bound} and Eq. \eqref{E:upper_bound_genK}, the upper bound depends only on delay values of uncongested links and it goes to zero for an ideal network (i.e. zero delay for uncongested links).

A notable point is that the size of a network (i.e., the number of links inside the network) grows with the number of boundary nodes. Therefore, for a fixed number of congested links, the number of uncongested links and the value of $\parallel \mathbf{x}_{S^c}\parallel_1$ increase as the number of boundary nodes increases. The upper bound of the derived estimation error depends upon the summation of the delay value of the uncongested links. In other words, the upper bound, for fixed $k$, is implicitly an increasing function of the number of boundary nodes. This explains the upward trend in Figure 5-(b).


\begin{figure}
\centering
\begin{tabular}{c}
\psfig{figure=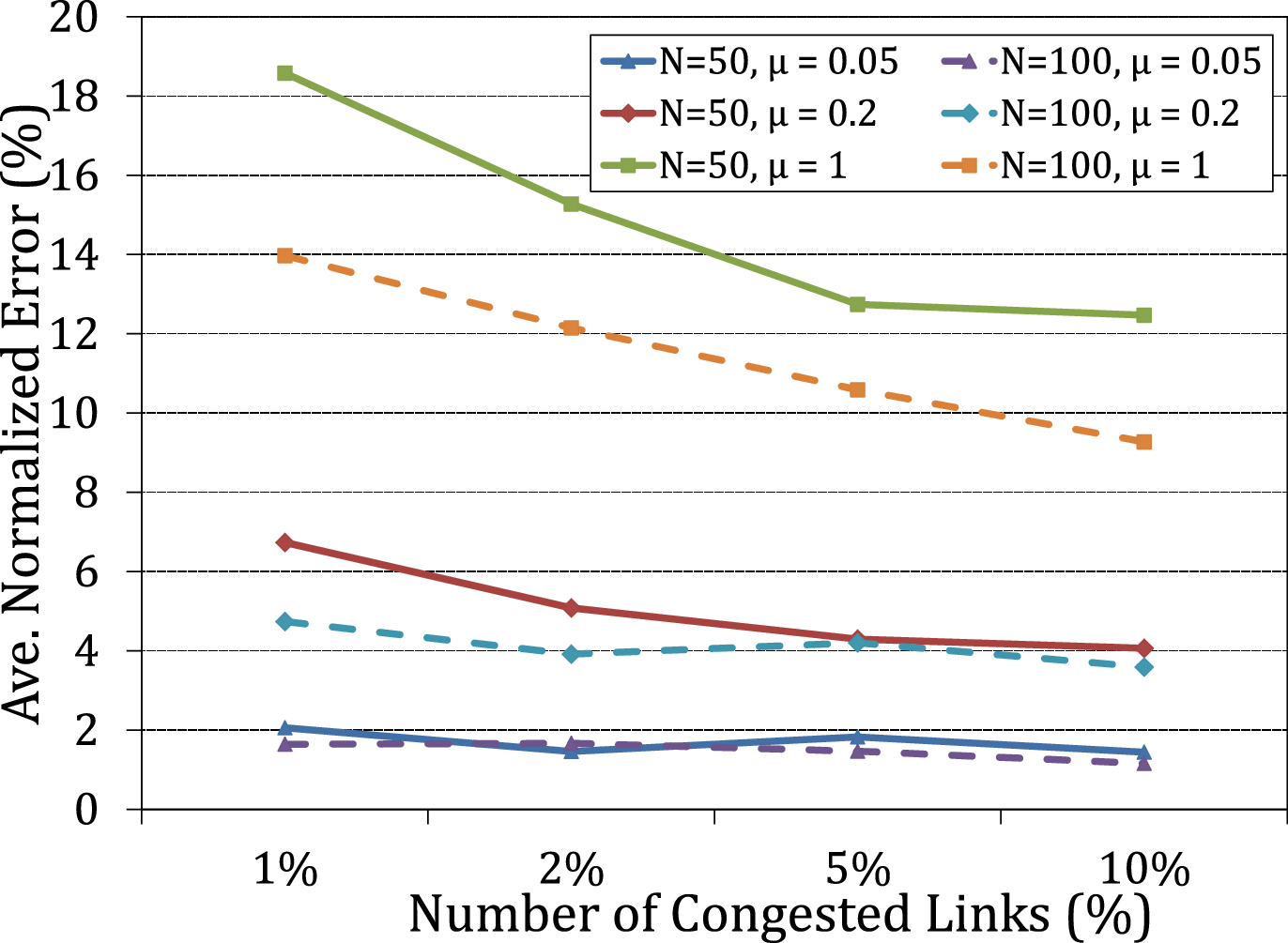,width=3.0in}
\end{tabular}
\caption{Average normalized estimation error in large networks with $N=50$ and $N=100$ boundary nodes. Networks satisfy conditions given in Definition \ref{Def:k-identifExpander} and $k\%$ of the links are deficient.}
\label{F:error_big}
\end{figure}

To evaluate the algorithm for large networks, we generate networks with 50 and 100 boundary nodes by implementing the process explained at the beginning of this section. For each number of boundary nodes, N = 50 and N = 100, 50 different network samples are generated. The average numbers of links in these samples are around 300 and 450 links, respectively. Figure \ref{F:error_big} presents the average normalized estimation error for delay recovery in $k$-identifiable expander networks, using the LP algorithm, where $k\%$ of the links are congested. The process of assigning delay to congested and uncongested links is similar to the previous scenario.

A notable point is that for a fixed value of $\mu$, the value of $\parallel\mathbf{x}\parallel_1$ increases as the number of congested links increases. Figure \ref{F:error_big} shows that the estimation error does not increase as fast as the value of $\parallel\mathbf{x}\parallel_1$. Hence, for a fixed $\mu$ value, therefore, the normalized estimation error exhibits a decreasing trend as the number of congested links increases; the trend occurs as the number of congested links satisfied the sparsity condition $k\ll |E|$. For the same reason, for a fixed $\mu$, the normalized estimation error for a large network, $N=100$, is slightly lower than the normalized estimation error with $N=50$ boundary nodes.

Finally, we evaluate the LP approach as applied to a detection (binary hypothesis testing) problem for $k$-identifiable expander networks. We study the effectiveness of LP optimization in classifying links as congested and uncongested for $k$-identifiable networks. To this end, we compare the delay caused by each link with a threshold. Links with delays greater than the threshold are categorized as congested, and vice versa. That is, we consider the simple detection rule:
\begin{equation}\label{E:detection}
\forall l\in E, \text{if }x_l\ge \tau \Rightarrow l\text{ is congested}
\end{equation}

\begin{figure}
\centering
\begin{tabular}{c}
\psfig{figure=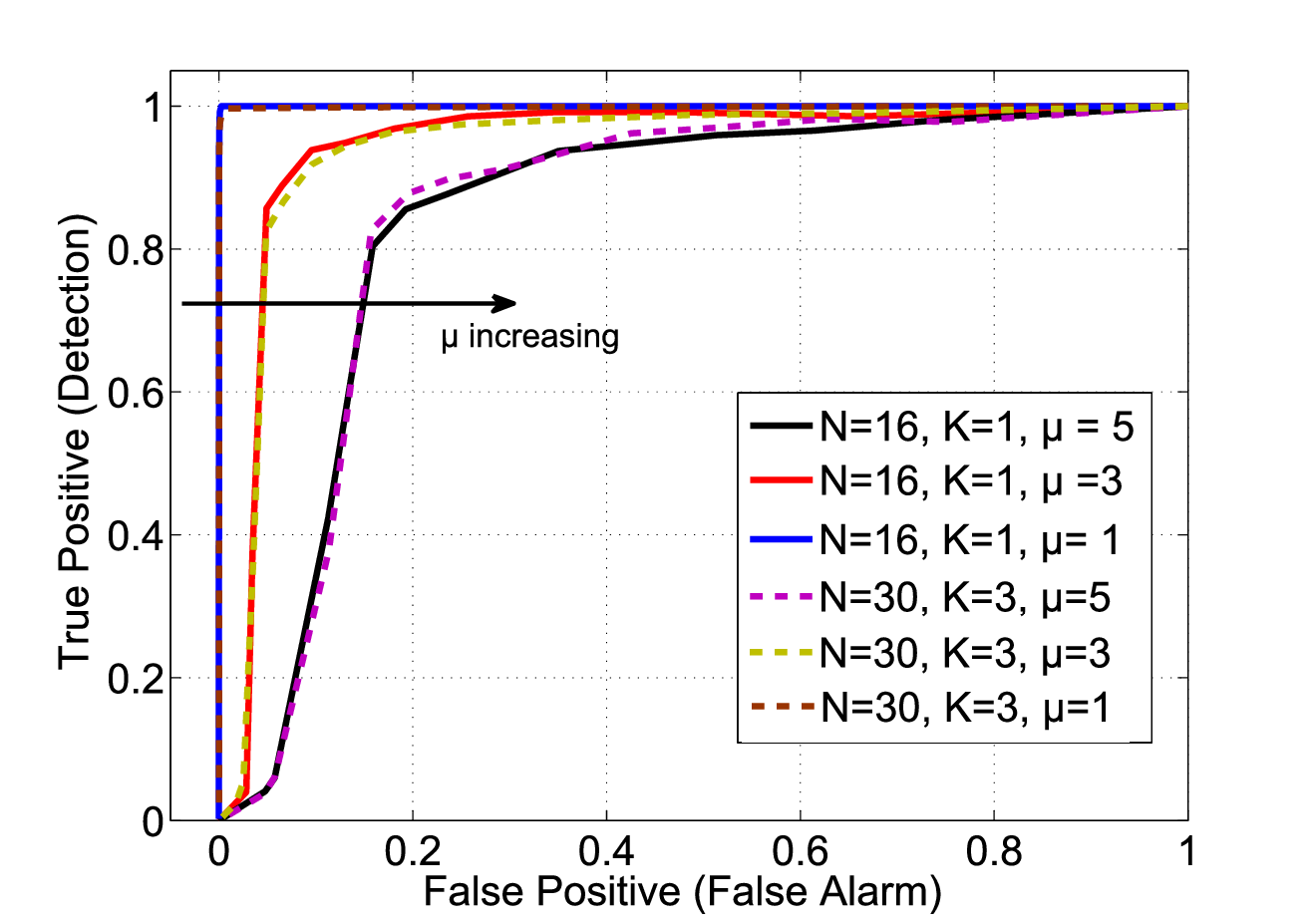,width=3.0in}\\
\end{tabular}
\caption{Receiver Operating Characteristic for the $k$-identifiable expander networks. When a link is uncongested, it experiences fixed delay value sampled from exponentially distributed delays given in Eq. \eqref{E:f_l}. To model congestion events, $k$ reference links are randomly selected and assigned a delay of $10$ ms.}
\label{F:ROC}
\end{figure}

Figure \ref{F:ROC} presents the receiver operating characteristic (ROC) for the above-mentioned detection rule, where $\tau$ changes from $ 0 \, \rightarrow \; \infty$. For $\mu=1$, the ROC plot is very close to that of the ideal receiver, indicating that the LP optimization in Eq. \eqref{E:LP} can distinguish between congested and uncongested links. An interesting observation from Figure \ref{F:ROC} is that detection performance depends only on the delay value of uncongested links and is independent of network size, number of congested links $k$, and delay value of congested links as long as $k$ satisfies the sparsity condition, i.e., $k\ll |E|$.

\subsection{Delay Estimation: Cumulative Distribution Function}

In this section, we compare our results with those produced by CF-estimator, one of the recent and state-of-the-art network delay estimators proposed in \cite{chen2010network}. In the study, the authors proposed a mixture model of characteristic functions for delay matrix $\mathbf{x}$ and developed a fast estimation algorithm based on generalized method of moments (GMM). The authors claim that this approach enables the use of more flexible models of heterogeneous network link delays, wherein the delays are non-discrete and may have different scales across all network links. We provide the cumulative probability of the normalized estimation error using both methods and show that LP optimization under the new proposed conditions achieves a lower error cumulative distribution.

Let $\mathbf{x}$ be the actual delay of the links and $\hat{\mathbf{x}}$ be the output of the delay estimator. Then we estimate empirically the following CDF:
\begin{equation}
P({{\parallel\mathbf{x}-\hat{\mathbf{x}}\parallel_1}\over {\parallel\mathbf{x}\parallel_1}} \le \delta).
\end{equation}

\begin{figure*}
\centering
\begin{tabular}{cc}
\psfig{figure=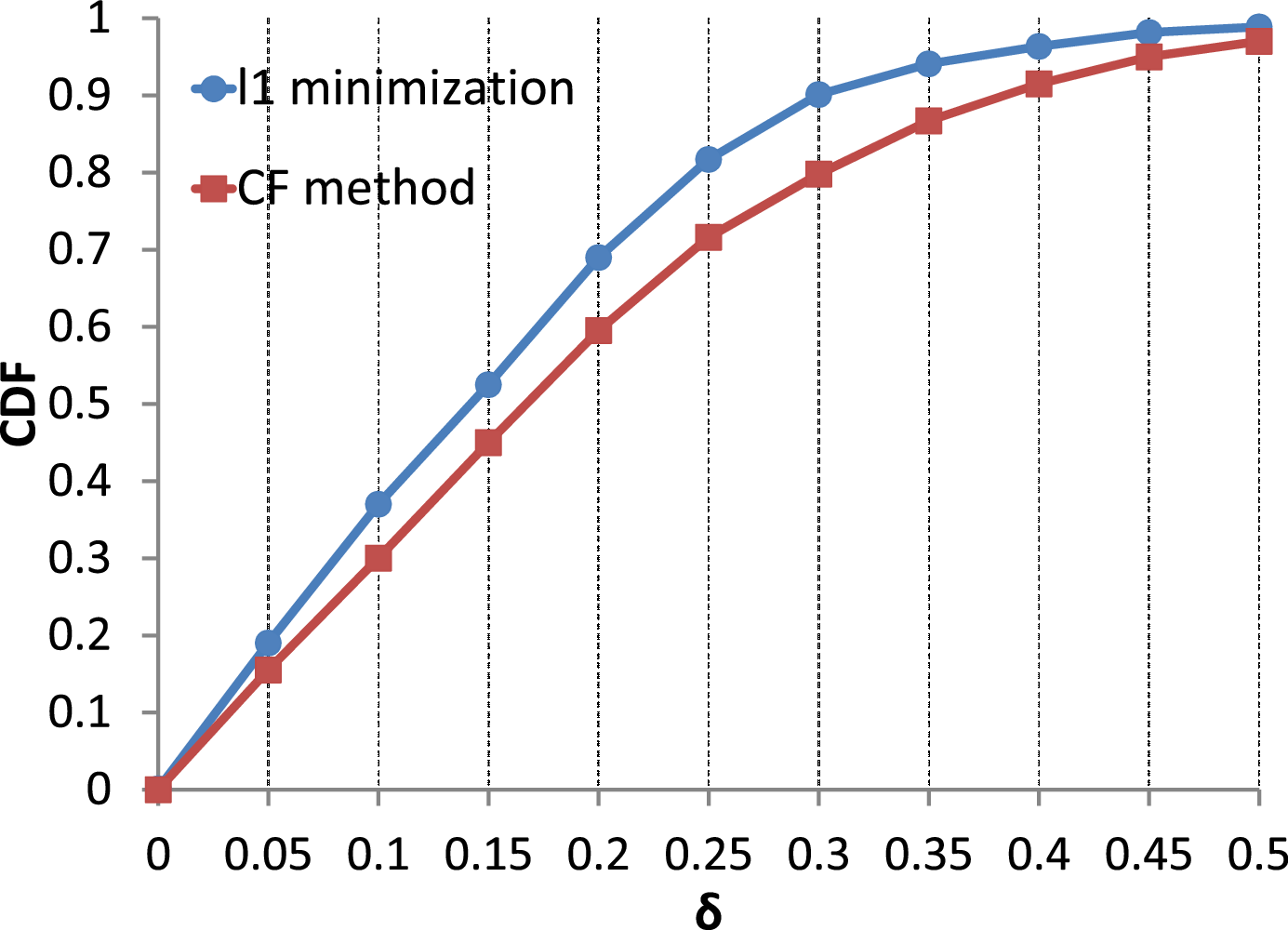,width=2.5in}
&
\psfig{figure=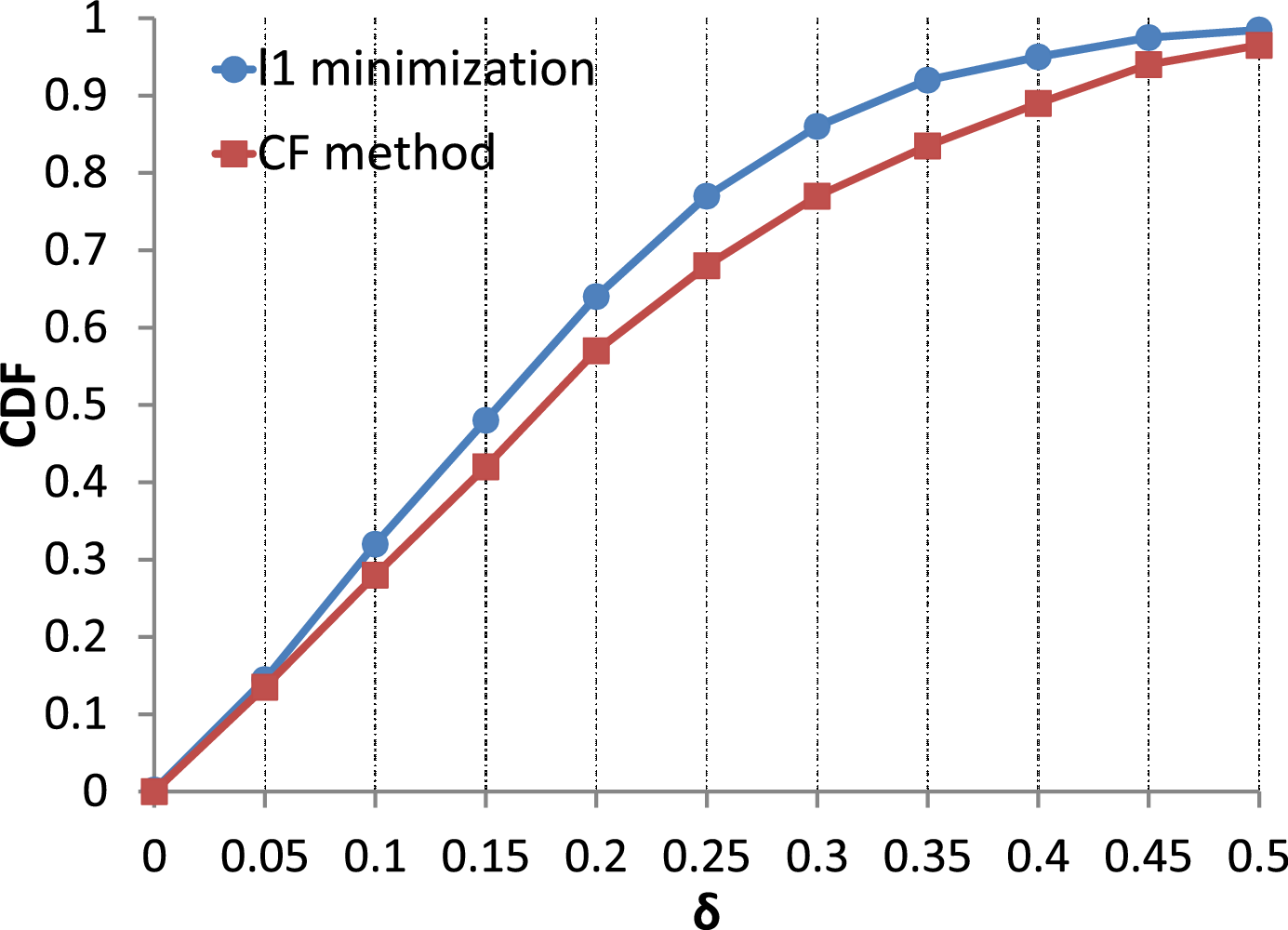,width=2.5in}
\\
(\textbf{a})
&
(\textbf{c})
\end{tabular}
\caption{Cumulative distribution function of estimation error for (a)$k=1$ (b)$k=3$. For $k$-identifiable expander networks (Definition \ref{Def:k-identifExpander}), the LP algorithm outperforms CF method.}
\label{F:CF_comparing}
\end{figure*}

Figure \ref{F:CF_comparing} presents the CDF of  the normalized estimation error when there are $k$ congested links inside the network for $k=1,3$ and $\delta \in [0,.5]$. For each CDF, 200 random networks are generated and for each generated network, link delays are assigned as described in Section \ref{S:delayEstimation} with $\mu\in\{0,.1,.2,1\}$. As can be seen, for $k$-identifiable networks, the LP algorithm outperforms CF algorithm, since it uses available sparsity as its side information.

\section{Conclusion}\label{S:conl}
In this work, we investigated the application of expander graphs and compressed sensing to network tomography. As shown by examples and simulation evidence, the current results on expander graphs do not apply to most of the networks. Hence, we modify some of the results to be more suitable for the delay estimation problem. We show that the number of Internet-topology-based networks satisfying new conditions is increased by 30\%. For those networks, we compare delay estimation based on compressed sensing with a state-of-the-art delay estimation algorithm.  The results show that compressed sensing provides better estimation, i.e., lower CDF of estimation error. Previous work has shown that in networks with large node degrees, recovery is typically better. Our work  supports this since networks with large node degrees have smaller $\epsilon$ for their corresponding expander graphs, resulting in smaller estimation error via $l_1$-optimization.

\bibliographystyle{IEEEtran}
\bibliography{IEEEabrv,overall.bib}

\appendices
\section{Proofs of Theorems}\label{SS:apx_finite}

We use the following convention to indicate the end of the proofs for theorems and lemmas. For main theorems and lemmas in the body of the manuscript, the end of the proof is denoted by $\small\blacksquare$; for preliminary lemmas that are needed to prove the main theorems,  end of the proof is denoted by $\tiny\square$.

Proof of \emph{Lemma} \ref{Tm:relax_epsilon}: to prove this lemma, we first need the following result which characterizes the null space of bi-adjacency matrix of an expander graph; it will be used to bound the error in the recovery of $\mathbf{x}$ from its compressed projection $\mathbf{y}$.

\begin{lemma}
Let $G(X,Y,H)$ be a $(2,d,\epsilon)$-expander with $\epsilon\le1/4$ and $\mathbf{A}_{m\times n}$ be its bi-adjacency matrix. Assume $\mathbf{w}$ lies in the null space of $\mathbf{A}$ (i.e., $\mathbf{Aw}=\mathbf{0}$) and let $S$ be any singleton set of coordinates of the $\mathbf{w}$, i.e., $ S = \{ i \}$, $i\in\{1,...,n\}$, representing the single large entry. Then,
\begin{equation}
\parallel \mathbf{w}_S\parallel_1\le 2\epsilon\parallel \mathbf{w}_{S^c}\parallel_1.
\label{E:w_s<w_s'}
\end{equation}
\label{Tm:ws<wsc}
\end{lemma}
\vspace{-8mm}
\emph{Proof}: let $\mathbf{A}^\prime$ be the submatrix of $\mathbf{A}$ containing rows from $N(S)$. Since $|S|=1$ and graph is left $d$-regular, $\parallel \mathbf{A}^\prime \mathbf{w}_S\parallel_1=\parallel \mathbf{A}\mathbf{w}_S\parallel_1=d \parallel \mathbf{w}_S\parallel_1$. We have
\begin{eqnarray}
0=\parallel \mathbf{A}^\prime\mathbf{w}\parallel_1&=&\parallel \mathbf{A}^\prime\mathbf{w}_S+\mathbf{A}^\prime\mathbf{w}_{S^c}\parallel_1 \\ \nonumber
                          &\ge& \parallel \mathbf{A}^\prime\mathbf{w}_S\parallel_1-\parallel \mathbf{A}^\prime\mathbf{w}_{S^c}\parallel_1 \\ \nonumber
                          &=& d\parallel \mathbf{w}_S\parallel_1-\parallel \mathbf{A}^\prime\mathbf{w}_{S^c}\parallel_1.
\end{eqnarray}

Each set of two nodes on the left has at least $2(1-\epsilon)d$ neighbor nodes on the right (via expansion). Since each node on the left has degree $d$, number of common nodes on the right-hand side is at most $2d-2(1-\epsilon)d = 2\epsilon d$. That means each column of $\mathbf{A}^\prime$ (except the one corresponding to $S$) has at most $2\epsilon d$ number of ones, yielding,
\begin{equation}
\parallel \mathbf{A}^\prime \mathbf{w}_{S^c}\parallel_1\le 2\epsilon d \parallel \mathbf{w}_{S^c}\parallel_1.
\end{equation}

Therefore, we have $\parallel \mathbf{w}_S\parallel_1\le 2\epsilon \parallel \mathbf{w}_{S^c}\parallel_1$, or equivalently
\begin{equation}\label{E:Null_extension}
\parallel \mathbf{w}_S\parallel_1 \le{2\epsilon \over {1+2\epsilon}}\parallel \mathbf{w}\parallel_1.
\end{equation}

\begin{flushright}
\tiny
$\square$
\end{flushright}
Consider $\mathbf{x}^\prime, \mathbf{x} \; \in \; \cal{N}(\mathbf{A})$, and let $\mathbf{y}=\mathbf{x}^\prime-\mathbf{x}$. Clearly $\mathbf{y}\in \cal{N}(\mathbf{A})$ and we have:
\begin{eqnarray}
\parallel \mathbf{x}\parallel_1&\ge& \parallel \mathbf{x}^\prime\parallel_1 \\ \nonumber
                                                &=&\parallel \mathbf{x}_S+\mathbf{y}_S\parallel_1+\parallel \mathbf{x}_{S^c}+\mathbf{y}_{S^c}\parallel_1 \\ \nonumber
                                                &\ge& \parallel \mathbf{x}_S\parallel_1-\parallel \mathbf{y}_S\parallel_1+\parallel \mathbf{y}_{S^c}\parallel_1 - \parallel \mathbf{x}_{S^c}\parallel_1  \\ \nonumber
                                                &=&\parallel \mathbf{x}\parallel_1-2\parallel \mathbf{x}_{S^c}\parallel_1+\parallel \mathbf{y}\parallel_1 - 2\parallel \mathbf{y}_S\parallel_1 \\ \nonumber
                                                &\ge& \parallel \mathbf{x}\parallel_1-2\parallel \mathbf{x}_{S^c}\parallel_1+(1-{4\epsilon \over {1+2\epsilon}})\parallel \mathbf{y} \parallel_1,
\end{eqnarray}
where in the last equality, Eq. \eqref{E:Null_extension} is used. Therefore we have:
\begin{equation}
\parallel \mathbf{x}^\prime-\mathbf{x} \parallel_1=\parallel \mathbf{y}\parallel_1\le {{2(1+2\epsilon)}\over {1-2\epsilon}}\parallel \mathbf{x}_{S^c} \parallel_1.
\end{equation}
\begin{flushright}
\small
$\blacksquare$
\end{flushright}

Proof of \emph{Theorem} \ref{Tm:expander_tomo}: with Lemma \ref{Tm:relax_epsilon} in place, Theorem 1 readily follows. Let $\mathbf{x}^\prime$ be the solution to the optimization problem in Eq. \eqref{E:LP}. That means $\mathbf{Rx}^\prime=\mathbf{Rx}$ and $\parallel \mathbf{x}^\prime\parallel_1\le \parallel \mathbf{x}\parallel_1$.
On the other hand, $G$ is a $(2,d,\epsilon)$-expander graph with the bi-adjacency matrix $\mathbf{R}$. Consequently, Eq. \eqref{E:xp_fe} holds for $\mathbf{x}$ and $\mathbf{x}^\prime$, and Theorem 1 is proved.
\begin{flushright}
\small
$\blacksquare$
\end{flushright}

Proof of \emph{Theorem} \ref{Tm:extensionExpander}: we prove the theorem for the case in which $G(X,Y,H)$ has only two expander subgraphs, since the general case follows using similar argument. Let $G_1(X_1,Y,H_1)$ with $|X_1|=m$, and $G_2(X_2,Y,H_2)$ with $|X_2|=n-m$, be two $d_i$-regular ($d_1\not= d_2$) subgraphs of $G(X,Y,H)$ with bi-adjacency matrices $\mathbf{R}_1$ and $\mathbf{R}_2$, respectively. Without loss of generality, we rename the elements in $X$ such that $\mathbf{R}=[\mathbf{R}_1\,\mathbf{R}_2]$.

Now, suppose two $1$-sparse vectors $\mathbf{u}$ and $\mathbf{v}$ have the same projection for matrix $\mathbf{R}_{r\times n}$, i.e., $\mathbf{Ru}=\mathbf{Rv}$. Without loss of generality, we assume $\parallel\mathbf{u}\parallel_1\ge\parallel\mathbf{v}\parallel_1$. Let $\mathbf{w}=\mathbf{v}-\mathbf{u}$. Clearly, $\mathbf{w}$ belongs to the null space of $\mathbf{R}$.  Moreover, partition $\mathbf{u}=[\mathbf{u}_1^t \; \mathbf{u}_2^t]^t$, $\mathbf{v}=[\mathbf{v}_1^t \;\mathbf{v}_2^t]^t$ and $\mathbf{w}=[\mathbf{w}_1^t \; \mathbf{w}_2^t]^t$. Then,
\begin{equation}\label{E:General_eq_wuv}
\mathbf{w}_1=\mathbf{v}_1-\mathbf{u}_1, \mathbf{w}_2=\mathbf{v}_2-\mathbf{u}_2.
\end{equation}

Let $S\subset \{1,2,\ldots,n\}$ be any singleton set of coordinates ($k=1$) of $\mathbf{w}$ and $\mathbf{R}^\prime$ is the submatrix of $\mathbf{R}$ corresponding to rows from $N(S)$. We consider the following two cases:

\textbf{Case 1:} $S\subset \{1,2,\ldots,m\}$:

In this case $S$ represents a node in $G_1(X_1,Y,H_1)$ which, by assumption, is a $(2,d_1,\epsilon)$-expander. Similar to the proof of Lemma 1, $\parallel\mathbf{R}^\prime \mathbf{w}_S\parallel_1=\parallel \mathbf{R}\mathbf{w}_S\parallel_1=d_1\parallel\mathbf{w}_S\parallel_1$. Thus,
\begin{eqnarray}\label{E:general_1}
0=\parallel\mathbf{R}^\prime\mathbf{w}\parallel_1&=&\parallel\mathbf{R}^\prime\mathbf{w}_S+\mathbf{R}^\prime\mathbf{w}_{S^c}\parallel_1\\ \nonumber
&\ge&d_1\parallel\mathbf{w}_S\parallel_1-\parallel\mathbf{R}^\prime\mathbf{w}_{S^c}\parallel_1.
\end{eqnarray}

Let $\mathbf{r}^t_i$ denote the $i$-th row of $\mathbf{R}^\prime$. The bipartite graph $G_1$ is left $d_1$-regular and hence matrix $\mathbf{R}^\prime$ has $d_1$ rows. Therefore, we derive an upper bound on $\parallel\mathbf{R}^\prime\mathbf{w}_{S^c}\parallel_1$ as below
\begin{eqnarray}\label{E:General_2}
\parallel\mathbf{R}^\prime\mathbf{w}_{S^c}\parallel_1 &=& \sum_{i=1}^{d_1} |\mathbf{r}^t_i\mathbf{w}_{S^c}| \\ \nonumber
&\stackrel{(1)}{\le}& \sum_{i=1}^{d_1}\sum_{j=1}^{n}r_{ij}|(w_{S^c})_j|\\ \nonumber
&=&\sum_{j=1}^{m}|(w_{S^c})_j| \sum_{i=1}^{d_1}r_{ij}+\!\!\sum_{j=m+1}^{n}\!\!\!\!|(w_{S^c})_j| \sum_{i=1}^{d_1}r_{ij},
\end{eqnarray}
where for inequality in the 1st step, we used the triangular inequality and the fact that $r_{ij}\in\{0,1\}$. Since $G_1(X_1,Y,H_1)$ is a $(2,d_1,\epsilon)$-expander, any two nodes on the left have at most $2\epsilon d_1$ neighbors on the right in common. That means, each column in $\mathbf{R}^\prime$ has at most $2\epsilon d_1$ nonzero entries  (i.e., $\sum_{i=1}^{d_1}r_{ij}\le 2\epsilon d_1$) for any $j\in\{1,2,\ldots,m\}\backslash S$. On the other hand, because $\mathbf{R}^\prime$ has $d_1$ rows, $\sum_{i=1}^{d_1}r_{ij}\le d_1$ for each $j\in\{m+1,\ldots,n\}$. Along with Eq. \eqref{E:General_2}, it results in:
\begin{eqnarray*}
\parallel\mathbf{R}^\prime\mathbf{w}_{S^c}\parallel_1\!\!&\le&\!\!2\epsilon d_1\sum_{j=1}^{m}|(w_{S^c})_j|+d_1\sum_{j=m+1}^{n}|(w_{S^c})_j|\\ \nonumber
\!\!&=&\!\!2\epsilon d_1\parallel\mathbf{w}_{1S^c}\parallel_1+d_1\parallel\mathbf{w}_{2}\parallel.
\end{eqnarray*}

After substituting above in Eq. \eqref{E:general_1} and rearranging the inequality, we have:
\begin{equation}\label{E:w_S_UPPER}
\parallel\mathbf{w}_S\parallel_1\le2\epsilon \parallel\mathbf{w}_{1S^c}\parallel_1+\parallel\mathbf{w}_{2}\parallel_1,
\end{equation}
which can be rewritten as:
\begin{equation}\label{E:w_S_1_UPPER_2}
\parallel\mathbf{w}_S\parallel_1\le{{2\epsilon}\over{1+2\epsilon}} \parallel\mathbf{w}_1\parallel_1+{1\over{1+2\epsilon}}\parallel\mathbf{w}_{2}\parallel_1.
\end{equation}

By assumption, $\parallel\mathbf{u}\parallel_1 \ge \parallel\mathbf{v}\parallel_1$. This further yields the following lower bound on $\mathbf{u}$:
\begin{eqnarray}\label{E:General_eq_1}
\parallel\mathbf{u}\parallel_1 \ge  \parallel\mathbf{v}\parallel_1 &=& \parallel \mathbf{u}_1+\mathbf{w}_1\parallel_1+ \parallel \mathbf{u}_2+\mathbf{w}_2\parallel_1 \\ \nonumber
&=&\parallel\mathbf{u}_{1S}+\mathbf{w}_{1S}\parallel_1+\parallel\mathbf{u}_{1S^c}+\mathbf{w}_{1S^c}\parallel_1+\\ \nonumber
&&\;\parallel \mathbf{u}_2+\mathbf{w}_2\parallel_1 \\ \nonumber
&\ge& \parallel\mathbf{u}_{1S}\parallel_1-(\parallel\mathbf{u}_{1S^c}\parallel_1+\parallel\mathbf{u}_2\parallel_1)+\\ \nonumber
&&\;(\parallel\mathbf{w}_{1S^c}\parallel_1+\parallel\mathbf{w}_2\parallel_1)-\parallel\mathbf{w}_{1S}\parallel_1
\end{eqnarray}

Because $S\subset \{1,\ldots,m\}$, we have $\parallel\mathbf{w}_2\parallel_1+\parallel\mathbf{w}_{1S^c}\parallel_1=\parallel\mathbf{w}_{S^c}\parallel_1$, $\parallel\mathbf{u}_2\parallel_1+\parallel\mathbf{u}_{1S^c}\parallel_1=\parallel\mathbf{u}_{S^c}\parallel_1$, and $\parallel\mathbf{w}_S\parallel_1=\parallel\mathbf{w}_{1S}\parallel_1$. So Eq. \eqref{E:General_eq_1} can be simplified as below:
\begin{eqnarray}\label{E:General_eq_2}
2\parallel\mathbf{u}_{S^c}\parallel_1 &\ge& \parallel\mathbf{w}_{S^c}\parallel_1-\parallel\mathbf{w}_{S}\parallel_1 \\ \nonumber
&=& \parallel\mathbf{w}\parallel_1-2\parallel\mathbf{w}_{S}\parallel_1.
\end{eqnarray}

By using Eq. \eqref{E:w_S_1_UPPER_2} we have:
\begin{equation}\label{E:u_Sc_1}
2\parallel\mathbf{u}_{S^c}\parallel_1\ge
{{1-2\epsilon}\over{1+2\epsilon}}\parallel\mathbf{w}_1\parallel_1-{{1-2\epsilon}\over{1+2\epsilon}}\parallel\mathbf{w}_2\parallel_1.
\end{equation}

By Eq. \eqref{E:General_eq_wuv} and triangular inequality we have:
\begin{equation}
\parallel\mathbf{w}_2\parallel_1\le\parallel\mathbf{u}_2\parallel_1+\parallel\mathbf{v}_2\parallel_1.
\end{equation}

Applying above inequality to Eq. \eqref{E:u_Sc_1} we have:
\begin{equation}\label{E:General_Upperound_Y}
{{1+2\epsilon}\over{1-2\epsilon}}\!\left[2\!\parallel\!\!\mathbf{u}_{S^c}\!\!\parallel_1\!
+{{2-4\epsilon}\over{1+2\epsilon}}\big(\!\!\parallel\!\!\mathbf{u}_{2}\!\!\parallel_1
+\parallel\!\!\mathbf{v}_{2}\!\!\parallel_1\big)\!
\right]\!\ge
\parallel\!\!\mathbf{w}\!\!\parallel_1\!,
\end{equation}
where we use the fact that $\parallel\mathbf{w}\parallel_1=\parallel\mathbf{w}_1\parallel_1+\parallel\mathbf{w}_2\parallel_1$. Clearly, $\parallel\mathbf{u}_{S^c}\parallel_1\ge\parallel\mathbf{u}_{2}\parallel_1$ and $\parallel\mathbf{v}_{S^c}\parallel_1\ge\parallel\mathbf{v}_{2}\parallel_1$. Therefore, the following inequalities hold:
\begin{equation}\label{E:General_Upperound_Y_simple}
{{4}\over{1-2\epsilon}}\parallel\mathbf{u}_{S^c}\parallel_1
+2\parallel\mathbf{v}_{S^c}\parallel_1
\ge
\parallel\mathbf{w}\parallel_1.
\end{equation}

Now let $j\in S^c$. There is a path $p^*$ which goes through either the link $j$ or the link in $S$ but not both (since it is a logical network). Let $\mathbf{r}^t_{p^*}$ be the corresponding row for $p^*$ in the routing matrix $\mathbf{R}$. Since $\mathbf{Ru}=\mathbf{Rv}$, we have $\mathbf{r}^t_{p^*}\mathbf{u}=\mathbf{r}^t_{p^*}\mathbf{v}$.

If $p^*$ only goes through link $j$ (and therefore not the link in $S$), the corresponding entry of the link in $S$ in $\mathbf{r}^t_{p^*}$ is zero. Hence we have $\parallel \mathbf{u}_{S^c}\parallel_1\ge \mathbf{r}^t_{p^*}\mathbf{u}$. Therefore, we can have the following upper bound for every entry of $v_j\,\,\forall j\in S^c$
\begin{equation}
\parallel\mathbf{u}_{S^c}\parallel_1\ge v_j\,\,\forall j\in S^c.
\end{equation}

Now, let assume that the $p^*$ goes through the link in $S$. By assumption $\parallel\mathbf{u}\parallel_1\ge\parallel\mathbf{v}\parallel_1$. Therefore,

\begin{equation}
\parallel\mathbf{u}\parallel_1-\mathbf{r}^t_{p^*}\mathbf{u}\ge\parallel\mathbf{v}\parallel_1-\mathbf{r}^t_{p^*}\mathbf{v}.
\end{equation}

Since $p^*$ does not go through the link $j$ we have $\parallel\mathbf{v}\parallel_1-\mathbf{r}^t_{p^*}\mathbf{v}\ge v_j$. On the other hand, because $p^*$ goes through the link $S$, its entry in $\parallel\mathbf{u}\parallel_1-\mathbf{r}^t_{p^*}\mathbf{u}$ is zero. Therefore, $\parallel\mathbf{u}_{S^c}\parallel_1\ge\parallel\mathbf{u}\parallel_1-\mathbf{r}^t_{p^*}\mathbf{u}$ and $\parallel\mathbf{u}_{S^c}\parallel_1\ge v_j$.

In summary, we always have $\parallel\mathbf{u}_{S^c}\parallel_1\ge v_j\,\, \forall j$. By adding up both side of the inequality for all $j\in S^c$ we have:
\begin{equation}
|S^c|\parallel\mathbf{u}_{S^c}\parallel_1\ge \parallel \mathbf{v}_{S^c}\parallel_1.
\end{equation}

Clearly  $n=|X|>|S^c|$. Therefore, the following upper bound is valid for $\mathbf{w}$:
\begin{equation}
({{4}\over{1-2\epsilon}}+2n)\parallel\mathbf{u}_{S^c}\parallel_1
\ge
\parallel\mathbf{w}\parallel_1.
\end{equation}

\textbf{Case 2:} $S\subset \{m+1,m+2,\ldots,n\}$:

By the same argument as Case 1, we have:
\begin{equation}\label{E:General_3}
0 = \parallel\mathbf{R}^\prime\mathbf{w}\parallel_1 \ge d_2\parallel\mathbf{w}_S\parallel_1-\parallel\mathbf{R}^\prime\mathbf{w}_{S^c}\parallel_1.
\end{equation}

As in case 1, we can put an upper bound on $\parallel\mathbf{R}^\prime\mathbf{w}_{S^c}\parallel_1$ as follows
\begin{eqnarray*}
\parallel\mathbf{R}^\prime\mathbf{w}_{S^c}\parallel_1\!\!&\le&\!\!d_2\sum_{j=1}^{m}|(w_{S^c})_j|+2\epsilon d_2\sum_{j=m+1}^{n}|(w_{S^c})_j|\\ \nonumber
\!\!&\le&\!\!d_2 \parallel\mathbf{w}_{1}\parallel_1+2\epsilon d_2\parallel\mathbf{w}_{2S^c}\parallel_1.
\end{eqnarray*}

Using above inequality and Eq. \eqref{E:General_3}, we have the following upper bound for $\parallel \mathbf{w}_S\parallel_1$.
\begin{equation}\label{E:w_S_UPPER_2nd}
\parallel\mathbf{w}_S\parallel_1\le2\epsilon \parallel\mathbf{w}_{2S^c}\parallel_1+\parallel\mathbf{w}_{1}\parallel_1,
\end{equation}

Finally, the following property is derived for vector $\mathbf{w}\in \mathcal{N}(\mathbf{R})$:
\begin{equation}\label{E:w_S_1_UPPER__2nd_2}
\parallel\mathbf{w}_S\parallel_1\le{{2\epsilon}\over{1+2\epsilon}} \parallel\mathbf{w}_2\parallel_1+{1\over{1+2\epsilon}}\parallel\mathbf{w}_{1}\parallel.
\end{equation}

By the same argument as case 1 we have:
\begin{equation}
({{4}\over{1-2\epsilon}}+2n)\parallel\mathbf{u}_{S^c}\parallel_1
\ge
\parallel\mathbf{w}\parallel_1.
\end{equation}

To prove Eq. \eqref{E:upper_bound} for Theorem 2, one can use above equation and follow the same argument as given in proofs of Lemma \ref{Tm:relax_epsilon} and Theorem \ref{Tm:expander_tomo}.

\begin{flushright}
\small
$\blacksquare$
\end{flushright}

Proof of \emph{Theorem} \ref{Tm:expander_tomo_regular_genK}: to prove  this theorem, we first prove Lemma \ref{Tm:errorOfLP_gen} which
characterizes the null space of $\mathbf{R}$.

\begin{lemma}\label{Tm:errorOfLP_gen}
Assume that $\mathbf{R}_{r\times n}$ is a bi-adjacency matrix of a $(2k,d, \epsilon)$-expander graph. Let $\mathbf{w}$ lie in the null space of $\mathbf{R}_{r\times n}$
(i.e., $\mathbf{Rw}=\mathbf{0}$) and let $\mathRV{S}$ be any set of coordinates of $\mathbf{w}$ with $|\mathRV{S}|\le k$, $ \mathRV{S} \subset \{1,...,n\}$. Then,
\begin{equation}
\mathbb{E}_{\mathRV{S}}[\parallel \mathbf{w}_{\mathRV{S}}\parallel_1]\le 2\epsilon\mathbb{E}_{\mathRV{S}}[\parallel\mathbf{w}_{\mathRV{S}^c}\parallel_1],
\label{E:y_Sy<Sc}
\end{equation}
where the expectation is taken over the selection of $\mathRV{S}$.
\end{lemma}

\emph{Proof}: in this lemma, we aim to show that for a vector $\mathbf{w}$ in null space $\mathbf{R}$, if we randomly select entries of $\mathbf{w}$ (up to $k$ entries), $l_1$-norm of those entries is smaller, on average, than $l_1$-norm of the rest of the vector.

Without loss of generality, we label the $n$ links inside the network $l_1,l_2,\ldots, l_n$ where the delay of link $l_j$ is the $j$th entry in delay vector $\mathbf{x}$. First, let fix $\mathRV{S}=S$. That is, $S$ is a realization of random set $\mathRV{S}$. Let $\mathbf{R}^\prime$ be the
submatrix of $\mathbf{R}$ that contains rows from $N(S)$. By definition of $\mathbf{w}_S$, we have $\parallel \mathbf{R}\mathbf{w}_S\parallel_1 = \parallel \mathbf{R}^\prime\mathbf{w}_S\parallel_1$. Thus,
\begin{eqnarray}\label{E:0>ws-ws'}
0=\parallel \mathbf{R}^\prime\mathbf{w}\parallel_1 &=& \parallel \mathbf{R}^\prime\mathbf{w}_S+\mathbf{R}^\prime\mathbf{w}_{S^c}\parallel_1 \\ \nonumber
                          &\ge& \parallel \mathbf{R}^\prime\mathbf{w}_S\parallel_1-\parallel \mathbf{R}^\prime\mathbf{w}_{S^c}\parallel_1.
\end{eqnarray}

\begin{figure*}
\centering
\begin{tabular}{ccc}
\psfig{figure=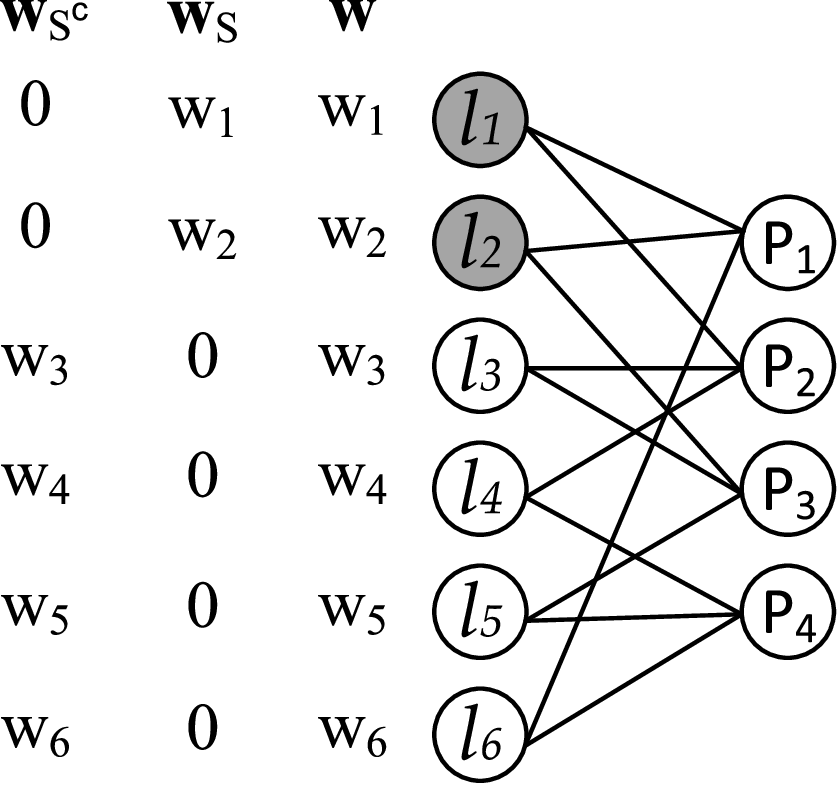,width=1.7in,rheight=1.25in}
,
&
$
\mathbf{R} =
\left[
\begin{array}{cccccc}
1&1&0&0&0&1\\
1&0&1&1&0&0\\
0&1&1&0&1&0\\
0&0&0&1&1&1\\
\end{array}
\right]
$
,
&
$\mathbf{R}^\prime =
\left[
\begin{array}{cccccc}
1&1&0&0&0&1\\
1&0&1&1&0&0\\
0&1&1&0&1&0\\
\end{array}
\right]
$
\end{tabular}
\caption{A realization of random set $\mathRV{S}$ for $S=\{1,2\}$ (congested links are $l_1$ and $l_2$) when $n=6$, $r=4$ and $d=2$. For the given $S$ we have $N(S)=\{P_1,P_2,P_3\}$.}
\end{figure*}

First, we impose a lower bound on $\parallel \mathbf{R}^\prime\mathbf{w}_{S}\parallel_1$. By using the result from Lemma 1 in \cite{berinde2008sparse},
we drive the following inequality:
\begin{equation}
\parallel \mathbf{R}^\prime\mathbf{w}_{S}\parallel_1\ge d(1-2\epsilon)\parallel \mathbf{w}_S\parallel_1.
\end{equation}

Substituting the above equation into Eq. \eqref{E:0>ws-ws'} yields:
\begin{equation}\label{E:R'Wsc>=d(1-e)Ws}
\parallel \mathbf{R}^\prime\mathbf{w}_{S^c}\parallel_1\ge d(1-2\epsilon)\parallel \mathbf{w}_S\parallel_1.
\end{equation}

Equality \eqref{E:R'Wsc>=d(1-e)Ws} is valid for every sample $S$ of random set $\mathRV{S}$. By taking the expectation over all possible outcomes of $\mathRV{S}$, which is also taken over all possible combinations of congested links, we have
\begin{equation}\label{E:E[0>ws-ws']}
\mathbb{E}_{\mathRV{S}}[\parallel \mathbf{R}^\prime\mathbf{w}_{\mathRV{S}^c}\parallel_1]\ge d(1-2\epsilon) \mathbb{E}_{\mathRV{S}}[\parallel \mathbf{w}_{\mathRV{S}}\parallel_1].
\end{equation}

Now, we aim to identify an upper bound for $\mathbb{E}_{\mathRV{S}}[\parallel\mathbf{R}^\prime \mathbf{w}_{\mathRV{S}^c}\parallel_1]$. Again, let fix $\mathRV{S}=S$. For a path $P$, define an indicator function $\mathRV{I}_{l\in P}$ as follows:
\begin{equation}
\mathRV{I}_{l\in P}=\left\{
                     \begin{array}{ll}
                       1, & \hbox{link $l$ belongs to path $P$;} \\
                       0, & \hbox{O.W.}
                     \end{array}
                   \right..
\end{equation}

The $d$-regularity assumption of the routing matrix $\mathbf{R}$ implies that the link $l$ belongs to exactly $d$ paths. For a link $l$, therefore, selecting a random path $P$ yields a probability of $\frac{d}{r}$ for the event that the path $P$ passes through $l$. Equivalently,
\begin{equation}
P(\mathRV{I}_{l\in P}=1) = \frac{d}{r} = \mathbb{E}[\mathRV{I}_{l\in P}].
\end{equation}

Recall that $\mathbf{R}^\prime$ is a submatrix of $\mathbf{R}$ that contains rows from $N(S)$. Thus, each row of $\mathbf{R}^\prime$ is a vector representation of a path. Let $\mathbf{r}^t_P$ be the row of $\mathbf{R}^\prime$ that corresponds to path $P$. Therefore,

\begin{eqnarray}
\parallel \mathbf{R}^\prime \mathbf{w}_{S^c}\parallel_1 &=& \sum_{P\in N(S)}\left| \mathbf{r}^t_P \mathbf{w}_{S^c}\right| \\ \nonumber
&\le& \sum_{P\in N(S)}\sum_{j\in S^c} |w_j|\mathRV{I}_{l_j\in P} \\ \nonumber
&=& \sum_{j\in S^c} |w_j|\sum_{P\in N(S)}\mathRV{I}_{l_j\in P} \\ \nonumber
&\le& \sum_{j=1}^n |w_j|\sum_{P\in N(S)}\mathRV{I}_{l_j\in P}.
\end{eqnarray}

The above-mentioned inequality is valid for every realization of the $S$ of random set $\mathRV{S}$. By taking the expectation over all possible outcomes of $\mathRV{S}$, which is also taken over all possible combinations of congested links, we have
\begin{eqnarray}\label{E:random_path_link}
\mathbb{E}_{\mathRV{S}}\left[\parallel \mathbf{R}^\prime \mathbf{w}_{\mathRV{S}^c}\parallel_1\right]&\le&\mathbb{E}_{\mathRV{S}}\left[\sum_{j=1}^n |w_j|\sum_{P\in N(\mathRV{S})}\mathRV{I}_{l_j\in P} \right] \\ \nonumber
&=& \sum_{j=1}^n |w_j|\mathbb{E}_{\mathRV{S}}\left[\sum_{P\in N(\mathRV{S})}\mathRV{I}_{l_j\in P} \right].
\end{eqnarray}

Random set $\mathRV{S}$ presents some random nodes on the left-hand side of the expander graph. These nodes are connected to set of nodes $N(\mathRV{S})$ on the right-hand side. Thus, $N(\mathRV{S})$ is also a random set that presents random nodes on the right. In the network, $N(\mathRV{S})$ represents random paths. Therefore, the last summation in Eq. \eqref{E:random_path_link} can be interpreted as follows: for a given index $j$, how many paths from the random paths $N(\mathRV{S})$ is the link $l_j$ connected to \footnote{Recall that we label the $n$ links inside the network $l_1,\ldots, l_n$ where the delay of link $l_j$ is the $j$th entry in delay vector $\mathbf{x}$.}. For any realization of $\mathRV{S}$, we have $|N(\mathRV{S})|\le kd$. In addition, $\mathRV{I}_{l_j\in P}$ are i.i.d random variables. Thus, for a given index $j$, link $l_j$ is connected, on average, to $\frac{k\;d^2}{r}$ number of nodes in $N(\mathRV{S})$. Therefore, we can derive the following upper bound:

\begin{equation}
\mathbb{E}_{\mathRV{S}}\left[\parallel \mathbf{R}^\prime \mathbf{w}_{\mathRV{S}^c}\parallel_1\right] \le \sum_{j=1}^n |w_j|kd\cdot \frac{d}{r} = \frac{kd^2}{r}\parallel\mathbf{w}\parallel_1.
\end{equation}

For $\epsilon>0$, $k\ll n$, $r\ll n$, in a $(2k,d,\epsilon)$-expander graph with $n$ number of nodes on the left and $r$ number of nodes on the right-hand side, the following inequality is proved \cite{capalbo2002randomness}:
\begin{equation}
\frac{2kd}{\epsilon}\le r.
\end{equation}

Finally, the following upper bound for $\mathbb{E}\left[\parallel \mathbf{R}^\prime \mathbf{w}_{S^c}\parallel_1\right]$ is derived:
\begin{eqnarray}\label{E:R'wSc<=}
\mathbb{E}_{\mathRV{S}}\left[\parallel \mathbf{R}^\prime \mathbf{w}_{\mathRV{S}^c}\parallel_1\right] &\le& \frac{d\epsilon}{2}\mathbb{E}_{\mathRV{S}}[\parallel\mathbf{w}\parallel_1] \\ \nonumber
&=& \frac{d\epsilon}{2}
\mathbb{E}_{\mathRV{S}}[\parallel\mathbf{w}_{\mathRV{S}}\parallel_1]+\frac{d\epsilon}{2}\mathbb{E}_{\mathRV{S}}[\parallel\mathbf{w}_{\mathRV{S}^c}\parallel_1].
\end{eqnarray}

Substituting Eq. \eqref{E:R'wSc<=} into Eq. \eqref{E:E[0>ws-ws']} results in
\begin{equation}
\mathbb{E}_{\mathRV{S}}[\parallel \mathbf{w}_{\mathRV{S}}\parallel_1]\le \frac{\epsilon}{(2-5\epsilon)}\mathbb{E}_{\mathRV{S}}[\parallel\mathbf{w}_{\mathRV{S}^c}\parallel].
\end{equation}

Given that $\epsilon\le \frac{1}{4}$, we have $\frac{\epsilon}{(2-5\epsilon)}\le 2\epsilon$ and therefore,
\begin{equation}\label{E:final_ES}
\mathbb{E}_{\mathRV{S}}[\parallel \mathbf{w}_{\mathRV{S}}\parallel_1]\le 2\epsilon\mathbb{E}_{\mathRV{S}}[\parallel\mathbf{w}_{\mathRV{S}^c}\parallel_1].
\end{equation}
\begin{flushright}
\tiny
$\square$
\end{flushright}
Inequality \eqref{E:final_ES} is similar to Eq. \eqref{E:w_s<w_s'}, except for the $\mathbb{E}_{\mathRV{S}}[\cdot]$. The rest of the proof is similar to that of Theorem \ref{Tm:expander_tomo}, in which $\parallel\mathbf{x}-\mathbf{x}^\prime\parallel_1$ is substituted with
$\mathbb{E}_{\mathRV{S}}[\parallel\mathbf{x}-\mathbf{x}^\prime\parallel_1]$.

\begin{flushright}
\small
$\blacksquare$
\end{flushright}

Proof of \emph{Theorem} \ref{Tm:expander_tomo_genK}: Eq. \eqref{E:upper_bound_genK}, the result of Theorem \ref{Tm:expander_tomo_regular_genK}, is similar to Eq. \eqref{E:upper_bound} in Theorem \ref{Tm:expander_tomo} by taking expectation over $\mathRV{S}$; i.e., $\mathbb{E}_{\mathRV{S}}[\cdot]$. To prove Theorem \ref{Tm:expander_tomo_regular_genK}, that is to relax the $d$-regularity condition, one can follow the exact same argument given in the proof of Theorem \ref{Tm:extensionExpander} and take the expectation over $\mathRV{S}$. Due to this similarity and for the sake of space saving, the proof is omitted here. 

\begin{flushright}
\small
$\blacksquare$
\end{flushright}

\end{document}